\documentclass[usenatbib]{mn2e}
\usepackage{graphicx}
\usepackage{float}
\usepackage{epsfig}
\usepackage{times}
\usepackage{subfigure}

\begin{document}

\title{Aperture Effects on Spectroscopic Galaxy Activity Classification}

\author[A. Maragkoudakis et al.]
{A. Maragkoudakis,$^1$ $^2$ A. Zezas,$^1$ $^2$ $^3$ M. L. N. Ashby,$^2$ S. P. Willner$^2$ \\
$^1$University of Crete, Department of Physics, Heraklion 71003, Greece \\
$^2$Harvard-Smithsonian Center for Astrophysics, Cambridge, MA 02138 \\
$^3$Foundation for Research and Technology - Hellas (FORTH), Heraklion 71003, Greece}

\maketitle
\begin{abstract}
Activity classification of galaxies based on long-slit and fiber spectroscopy can be strongly influenced by aperture effects. Here we investigate how activity classification for 14 nearby galaxies depends on the proportion of the host galaxy's light that is included in the aperture. We use both observed long-slit spectra and simulated elliptical-aperture spectra of different sizes. The degree of change varies with galaxy morphology and nuclear activity type. Starlight removal techniques can mitigate but not remove the effect of host galaxy contamination in the nuclear aperture. Galaxies with extra-nuclear star formation can show higher [O\,{\sc iii}]~$\lambda$5007/H$\beta$ ratios with increasing aperture, in contrast to the naive expectation that integrated light will only dilute the nuclear emission lines. We calculate the mean dispersion for the diagnostic line ratios used in the standard BPT diagrams with respect to the central aperture of spectral extraction to obtain an estimate of the uncertainties resulting from aperture effects.
\end{abstract}

\section{INTRODUCTION}

Energy production in galaxies is based on two major processes, sometimes mutually contributing to the total energy output. Star formation is the most common process, taking place throughout galaxies, while accretion of matter onto a supermassive black hole in the nucleus is another, less frequent, mechanism of energy production. The energetic nuclear regions associated with black hole accretion, referred to as Active Galactic Nuclei (AGN), are confined to the centers of galaxies. Understanding these processes is key to understanding galaxy evolution, including black hole growth in the center of galaxies, star formation, and possible correlations between them. Activity classification diagnostics provide tools to reveal the dominant energy production mechanism, signifying either processes of star-formation throughout the galaxy, or accretion in the nucleus. However, the distance dependence of physical size on angular size introduces observational difficulties (including starlight contamination from the host galaxy), making AGN identification a difficult and sometimes complex task. An accurate activity classification scheme that accounts for these complexities is essential to understand energy production in galaxies and subsequently galaxy evolution.

Visible emission-line ratio diagnostics are probably the most widely used and best-calibrated method of galaxy activity classification. Baldwin, Phillips and Terlevich (1981), hereafter BPT, and later Veilleux and Osterbrock \citeyearpar{VO87}, have shown that emission-line galaxies can be distinguished as star-forming, Seyfert, or Low-ionization nuclear emission-line regions \citep{Heckman80} using the intensity ratios of two pairs of emission lines. Three standard diagnostic diagrams are used, based on [O\,{\sc iii}]~$\lambda$5007/H$\beta$, [N\,{\sc ii}]~$\lambda$6583/H$\alpha$, [S\,{\sc ii}]~$\lambda\lambda$6716, 6731/H$\alpha$, or [O\,{\sc i}]~$\lambda$6300/H$\alpha$. Because the relative intensity of spectral lines of different excitation energy depends on the hardness of the ionizing continuum, galaxies with different sources of ionizing radiation occupy different locations in the diagrams. Kewley et al. \citeyearpar{Ke01} defined a theoretical upper limit on the location of star-forming galaxies in the BPT diagrams using H {\sc ii} region model spectra. Later Kauffmann et al. \citeyearpar{Kauff03} defined an empirical line delineating the pure star-forming galaxies in the [O\,{\sc iii}]$\lambda$5007/H$\beta$ versus [N\,{\sc ii}]~$\lambda$6583/H$\alpha$ diagnostic diagram based on a sample of 122,808 galaxies from the Sloan Digital Sky Survey \citep{SDSS}. In between the empirical pure starburst and the theoretical maximum starburst lines lie the composite or transition objects (TO), introduced by Ho, Filippenko \& Sargent \citeyearpar{Ho93}, which may have both star-burst and AGN activities. A refinement of the BPT diagrams was introduced by Kewley et al. \citeyearpar{Ke06}, who calculated the distinguishing line that separates Seyferts and LINERs on the standard optical diagnostic diagrams of [S\,{\sc ii}]~$\lambda\lambda$6716, 6731/H$\alpha$ and [O\,{\sc i}]~$\lambda$6300/H$\alpha$.

The activity classification obtained with visible-light diagnostic diagrams should be treated with caution. In general, the emission-line spectra of galaxy nuclei will be contaminated to some degree by unrelated stellar light that falls in the aperture; the problem is worse at high redshifts. In early-type galaxies in particular, where evolved stars dominate the light from galaxy bulges, absorption features in the stellar spectra will affect the strength of most emission lines. A demonstration of this effect was presented by Moran et al. \citeyearpar{Moran02} for distant galaxies, where their angular diameters were comparable to the widths of the slits used to measure their spectrum. Moran et al. obtained integrated and nuclear spectra of nearby Seyfert 2 galaxies, known to be absorbed X-ray sources, and showed that in their integrated spectra the emission lines were far less prominent or even completely obliterated compared to their respective nuclear spectra. These effects could result in a different activity classification of a galaxy depending on the region used to extract its spectrum. The effect becomes very important in the case of SDSS spectra of distant galaxies, where the 3$\arcsec$ diameter fibers used by SDSS may include a large portion of host galaxy's starlight or in some cases the whole galaxy's light. It is important to address this issue and examine if starlight subtraction methods are capable of correcting it.

Over the last decade starlight subtraction methods that can estimate and subtract the contamination by the host galaxy have been introduced (e.g., \citealp{Sarzi07}; \citealp{Hao05}). One of the most common approaches for starlight subtraction involves the use of templates, derived from spectra of several different galaxies (e.g., \citealp{Ho97a}). More recent methods (e.g., STARLIGHT; \citealp{Starlight1, Starlight2}) fit a model spectrum constructed from population synthesis techniques using a linear combination of stellar libraries. However, due to multiple parameters involved in the fitting process (e.g., metallicity, multiple stellar populations, extinction) along with the complexity of dealing with the AGN component, star formation, and the interstellar medium (ISM), the degree to which starlight subtraction methods can account for the contribution of starlight is still unclear.

Even with optimal starlight subtraction, emission line ratios can still vary with aperture size. One of the major reasons is the presence of extra-nuclear star-forming activity enhancing the integrated flux of emission lines. Another factor can be the metallicity gradient observed in disk galaxies (e.g., \citealp{Ferguson}), as lower-metallicity H {\sc ii} regions in the outer parts of galaxy discs are capable of producing high-excitation emission lines. When observing the integrated spectra of galaxies both effects may impact the line ratios and subsequently galaxy classification.

This paper examines how the BPT activity classification changes as a function of how much of the host galaxy's light that is included in the extraction aperture and the degree to which starlight removal techniques can mitigate the effect of host galaxy light. We use long-slit spectra for different types of galaxies in order to extract spectra of different-sized apertures and measure how the activity classification depends on the amount of host galaxy light included. Section 2 presents the study sample and reviews its properties. Section 3 describes the spectral extraction techniques used and the analysis thereof. Section 4 summarizes our results and discusses the effectiveness of starlight subtraction, and Section 5 presents our conclusions and discusses how spectra of more distant galaxies are affected by aperture effects.

\section{SAMPLE SELECTION}

The galaxies used in our analysis were selected from the Star Formation Reference Survey \citep{SFRS11}, an on-going multi-wavelength project studying the properties of star formation in nearby galaxies. Nearby galaxies with large surface are optimal candidates to test aperture effects by extracting spectra from incrementally increasing aperture widths. Therefore, for this study we excluded point-like sources, while at the same time we avoided galaxies with major axis diameter similar to the spectrograph slit length. The last criterion was introduced in order to avoid sky-subtraction procedures which would require a separate sky exposure, keeping the analysis as simple and unbiased as possible. Because we are examining how activity classification is affected by aperture effects, we included galaxies of different activity classifications, (star-forming, LINER, Seyfert) based on their nuclear spectra (Maragkoudakis et al, in prep). Out of the 50 objects observed so far, 11 fulfill the above criteria. 

In order to account for different morphological types the SFRS sample was supplemented by elliptical galaxies, known to host an AGN. The galaxies used were 3C~033, 3C~084, and 3C~296, selected based on the sample of \cite{Buttiglione}. The basic parameters of the final sample of 14 galaxies used in our study are presented in Table \ref{table1}.

With the exception of NGC~4704 at 122.8 Mpc, UGC~9412 at 198.7 Mpc, and 3C~033 at 252 Mpc, the objects in the sample are located within 92 Mpc with a mean distance of 79.8 Mpc. The majority of the galaxies are of spiral morphological types based on NASA/IPAC Extragalactic Database (NED), which is based on de Vaucouleurs system \citep{DV1959}. Their integrated photometry was obtained from the SDSS database and corresponds to G-band Petrosian magnitudes \citep{Blanton01}.

\begin{table*}
\begin{minipage}{144mm}
\caption{Galaxy Characteristics.}
\label{table1}
\begin{tabular}{@{}lcccccc}
 \hline
Galaxy & Morphology & SDSS $m_{g}$ & Distance & Exposure & Classification & D25 Angular Diameter \\
 & & (mag) & (Mpc) & (s) & (Nucleus) & (arcmin) \\
 \hline
NGC 2608 & SB(s)b & 13.520 $\pm$ 0.021 & 36.3 & 1800 & H\,{\sc ii} & 2.3 (1.7) \\
NGC 2731 & S? & 13.756 $\pm$ 0.081 & 35.0 & 900 & H\,{\sc ii} & 0.8	(0.9)\\
NGC 3306 & SB(s)m? & 14.025 $\pm$ 0.007 & 46.6 & 1800 & H\,{\sc ii} & 1.3 (0.9) \\
NGC 4412 & SB(r)b? pec & 12.972 $\pm$ 0.002 & 30.6 & 1200 & Sy2 & 1.4 (1.1)\\
NGC 4491 & SB(s)a & 13.327 $\pm$ 0.003 & 16.8 & 600 & H\,{\sc ii} & 1.7 (1.6) \\
NGC 4500 & SB(s)a & 13.070 $\pm$ 0.037 & 52.0 & 600 & H\,{\sc ii} & 1.6 (0.9) \\
NGC 4704 & SB(rs)bc pec & 14.255 $\pm$ 0.004 & 122.8 & 1200 & Sy2 & 1.0 (0.9) \\
NGC 4868 & SAab? & 13.662 $\pm$ 0.109 & 74.0 & 600 & H\,{\sc ii} & 1.6 (1.4) \\
NGC 5660 & SAB(rs)c & 13.358 $\pm$ 0.022 & 38.9 & 1800 & H\,{\sc ii} & 2.8 (1.6)\\
UGC 6732 & S0? & 13.533 $\pm$ 0.002 & 53.6 & 600 & LINER & 0.9 (1.6) \\
UGC 9412 & S? & 13.913 $\pm$ 0.003 & 198.7 & 1800 & Sy1 & 0.6 (0.4) \\
3C 033 & -	& 15.772 $\pm$ 0.005 & 251.8 & 1800 & Sy2 & 0.7 (0.8) \\
3C 084 / NGC 1275 & pec & 12.079 $\pm$ 0.659 & 68.2 & 600 & LINER & 2.2 (2.0) \\
3C 296 / NGC 5532 & - & 12.698 $\pm$ 0.009 & 91.2 & 900 & Sy2 & 0.6 (0.7) \\
 \hline
\end{tabular}

\medskip
Column (1) Galaxy name; Column (2) Morphological types obtained from NED; Column (3) Integrated SDSS G-band apparent magnitudes taken from Ashby et al. \citeyearpar{SFRS11}; Column (4) Distances in Mpc taken from Ashby et al. for the SFRS sample and NED for the elliptical galaxies; Column (5) Exposure time of each object's observation in seconds; Column (6) Activity classifications based on the nuclear spectra of galaxies (Maragkoudakis et al, in prep); Column (7) D25 angular diameters taken from Ashby et al. \citeyearpar{SFRS11} for the SFRS galaxies, and from the Third Reference Catalogue of Bright Galaxies (RC3; \citet{DV1991}) for the elliptical galaxies except for 3C033, which was obtained from 2MASS $K_{s}$ band survey (\cite{2MASS}). The value in the parentheses is the maximum angular diameter used for spectral extraction.
\end{minipage}
\end{table*}

\section{OBSERVATIONS AND ANALYSIS}

The long-slit spectra were acquired using the FAST Spectrograph \citep{FAST} mounted at the 60-inch Tillinghast telescope at Fred Lawrence Whipple Observatory during the period from March to May 2012. We used a $3\arcsec$ wide, $6\arcmin$ long slit oriented along the major axis of each galaxy with a 300 l/mm grating providing a dispersion of 1.47 \AA\ pixel$ ^{-1} $, and resolution of 322 km/s (5.993 \AA) as measured from the full width at half maximum (FWHM) of the O\,{\sc i} 5577 \AA\ sky line. The exposure times varied from 600 to 1800 seconds, and the spectra were obtained using the 2688 $\times$ 512 pixel FAST3 CCD, covering the 3500-7400 \AA\ spectral range. 

The spectra were analyzed using IRAF (Image Reduction and Analysis Facility; Tody, D. \citeyear{IRAF}). All standard initial procedures of bias subtraction and flat fielding as well as wavelength calibration using arc-lamp exposures, were performed on the two-dimensional CCD images. The integrated spectra of all galaxies were extracted at first using the ``APALL" task. Extracting the integrated spectra is an essential part of the analysis in order to obtain their activity classification based on the standard long-slit spectroscopic analysis and also to estimate the full extent of the galaxies along the slit. The size of the extraction aperture was based on the width of the spatial profile of the spectrum and was defined up to the point that the galaxy's luminosity was indistinguishable from the level of the background. The sky subtraction was also performed by ``APALL", and the level of the background was defined from regions in the spatial direction of the spectrum that do not contain any light from the source. The uncertainties were also calculated by ``APALL" as the square root of the counts in each dispersion-axis bin. Flux calibration was applied using standard star exposures, which were processed in the same way as the galaxy spectra. Lastly, we measured the location of characteristic lines from all galaxy spectra in order to calculate their redshift and correct them to their rest frame. 

Our first approach is to examine line ratios along the slits to see how the line ratios change and possibly influence classifications as a larger portion of starlight is gradually incorporated in the extracted spectrum. To achieve this, we began the spectral extraction from a small central aperture (3 pixels (3.51\arcsec) wide) and repeated the process by increasing the aperture width in steps of 2 pixels (2.34\arcsec) along the major axis until the entire galaxy was covered. The minimum size of the central aperture was determined from the average value of seeing at the time of the observations. We obtained 10 to 42 sub-spectra for each galaxy depending on its apparent size. All the steps of flux calibration and redshift correction were also performed in each individual sub-spectrum of every galaxy before proceeding to measure their emission lines.

\begin{figure}
\begin{center}
\includegraphics[keepaspectratio=true, scale=.46]{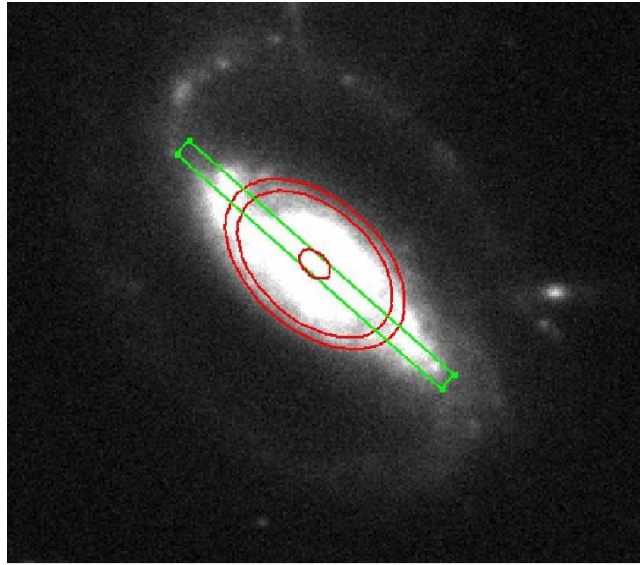}
\caption{SDSS $g$-band image of NGC~4500. Parallel lines show the long slit, and ellipses show an example annulus, in this case 34$\arcsec$ in diameter. The intersection of the long slit and the annulus apertures denotes the rectangular regions used to create the spectrum of the elliptical annulus.}
\label{apertures}
\end{center}
\end{figure}

The next step in our analysis was to simulate elliptical apertures of various sizes based on the spectra extracted from different apertures along the major axis. This way we produce spectra resembling those obtained with optical fiber spectrographs. To do so we started with a central (3-5 pixels wide along the major axis) aperture. In order to simulate spectra from larger elliptical apertures we incrementally added spectra from emulated elliptical annuli of increasing radius. Each successive annulus was derived from the spectrum of two 2-pixel wide (2.34$\arcsec$) rectangular apertures at equal distance on either side from the center of the galaxy. Then, using G-band images of the galaxies taken from the SDSS DR8 \citep{SDSSDR8}, we measured the broadband emission in each of the rectangular apertures as well as in the elliptical annulus with major and minor axes enclosing these regions and position angle equal to that of the long-slit (Figure \ref{apertures}). The sky background was defined from source-free regions on the SDSS images and then subtracted from the regions of interest. The ratio of the intensity in the elliptical annulus to the intensity inside the rectangular apertures provides the multiplicative factor needed to simulate the spectrum of an elliptical annulus based on the spectrum of the rectangular apertures. This way we emulate the spectrum from an elliptical annulus, assuming that the spectrum along the major axis is representative of the spectrum in the rest of the galaxy at the same galactocentric radius. This is an important step in order to account for the increasing area of the galaxy in larger galactocentric radii that is not accounted for in the equal rectangular apertures along the long-slit spectra.

For the central region of each galaxy, we used a single rectangular region to extract the spectrum and an elliptical region around it to calculate the weight and thus create a central elliptical spectrum, similarly to the process described previously. Adding consecutive simulated elliptical annuli spectra gave us the simulated spectra of elliptical apertures of varying sizes. 

In order to investigate the efficiency of starlight subtraction and the degree to which it can improve the inclusion of additional extra-nuclear light, we measured emission lines from simulated spectra with and without starlight subtraction. We removed the starlight contribution from all the long-slit and integrated elliptical aperture sub-spectra of each galaxy using the STARLIGHT v.04 starlight subtraction code. STARLIGHT fits the observed spectrum using linear combinations of simple stellar populations (SSPs) based on the models of Bruzual \& Charlot (2003). The fit was performed on emission-line-free regions of the spectrum. The best-fit combination was subtracted, and reddening corrections were performed. The extinction ($A_{V}$) is a fitted parameter applied to the unreddened stellar models. In our analysis we used the Cardelli reddening law \citep{CCM89} with $R_v$ = 3.1, defined as the ratio of total to selective extinction: $R_v \equiv A_v/E(B-V) $. To fit the starlight components we used a base of 138 SSPs, consisting of populations of 23 ages between 1Myr and 13Gyr and 6 metallicities ranging from 0.005 to 2.5 $Z\odot$.

After starlight subtraction we measured the intensities of the forbidden lines [O\,{\sc iii}]~$\lambda$5007, [N\,{\sc ii}]~$\lambda$6583, [S\,{\sc ii}]~$\lambda\lambda$6716, 6731, [O\,{\sc i}]~$\lambda$6300/H$\alpha$, and the Balmer H$\beta$ and H$\alpha$ lines for all the galaxies in both cases of long-slit and simulated-elliptical spectra. In the case of the simulated elliptical spectra we also measured emission lines without applying starlight subtraction. For this task we used SHERPA \citep{SHERPA}, a modeling and fitting application designed for CIAO, the Chandra (X-Ray Observatory) Interactive Analysis of Observations software package \citep{CIAO}. One of the advantages of SHERPA is that we are able to fit complex multi-component models and obtain the uncertainties on all parameters in the fit. The uncertainties were calculated (within 1$\sigma$ confidence intervals) by varying a given parameter's value along a grid of values while at the same time the values of all other model parameters were allowed to float to new best-fit values. We measured the emission line intensities by fitting them with a set of Gaussian functions and a constant for the continuum. All three characteristic parameters of each Gaussian (center, intensity, full width at half maximum (FWHM)) were generally free to vary. In many cases, linking parameters such as the FWHM of emission line doublets was required and applied individually in each set of lines. Additionally, in certain cases a broad component was required along with the narrow one in order to fit the Balmer emission lines. Such components can be the result of galaxy rotation or, in the case of Seyfert 1 galaxies, of emission from the broad-line region (BLR), producing line widths up to $10^{4}$ km/s.
 
The resulting diagrams are presented in Figures \ref{BPT1}a - \ref{BPT4}d. In NGC~4491 and UGC~6732 we could not measure any emission lines without applying starlight subtraction because they are lost in the stellar absorption features. For both NGC~4491 and NGC~4868, even in the case of starlight subtracted spectra, we were only able to obtain emission line measurements out to approximately half of their semi-major axis. NGC~4704 has a non starlight-subtracted ``path" on the BPT diagrams starting at the point where the diagnostic lines were measurable, approximately at half semi-major axis distance, and extended to its final largest aperture.

The light encompassed in a given aperture depends on radius. To calculate flux as a function of radius and thus the half-light radius, we built a curve of growth from the SDSS $g$-band images. Additionally, we plot on the same graphs the H$\alpha$ and H$\beta$ emission line equivalent widths measured from the non-starlight subtracted spectra in order to have a sense on the continuum change and its impact on the emission lines due to stellar Balmer absorption lines. The results are presented in Figures \ref{6PLOTS1}a - \ref{6PLOTS4}d. In most cases all line ratios tend to vary in the same way as a function of galactocentric radius. However, the range of their  actual values differs as a result of  the  different sources of ionizing photons when moving from the nucleus  to the outer parts of galaxies. There are also cases (e.g. NGC~2608, NGC~4500)  where the equivalent widths of H$\alpha$ and H$\beta$ as a function of galactocentric radius or encompassed flux appear to show the opposite behavior compared to the BPT line-ratio. This is a good indication that the Balmer lines are those that are driving the trends of the diagnostic line ratios and hence the resulting ``path" on the diagrams.

Based on the previous analysis we calculated the dispersion of every diagnostic line ratio for the H\,{\sc ii} and Seyfert galaxies. This was derived from their individual elliptical sub-spectra with respect to their central aperture. Then, we calculated the mean of the above dispersion for each galaxy class with its corresponding standard deviation. In this way we obtained a sense of the uncertainties of the line-ratios used in the BPT diagrams due to the effect of the host galaxy light, which is particularly useful for higher-Z galaxies that are generally unresolved. As we see in Table \ref{table2}, the largest scatter is observed in the [O\,{\sc iii}]~$\lambda$5007/H$\beta$ ratio of Seyfert galaxies. This is expected as the higher ionization [O\,{\sc iii}]~$\lambda$5007 line drops abruptly as we move farther out from the center of galaxies with AGN, as discussed in Section 4.

\begin{table}
\begin{center}
\caption{Diagnostic line ratio dispersion.}
\label{table2}
\begin{tabular}{@{}lcccc}
 \hline
 & \multicolumn{2}{c}{H\,{\sc ii}} & \multicolumn{2}{c}{Seyfert} \\
 \hline
\multicolumn{1}{c}{Line-Ratio} & $\mu_{\sigma}$ & $\sigma$ & $\mu_{\sigma}$ & $\sigma$\\
 \hline
log([O\,{\sc iii}]~$\lambda$5007/H$\beta$) & 0.21 & 0.16 & 0.45 & 0.49\\
log([N\,{\sc ii}]~$\lambda$6583/H$\alpha$) & 0.18 & 0.08 & 0.34 & 0.31\\
log([S\,{\sc ii}]~$\lambda\lambda$6716, 6731/H$\alpha$) & 0.16 & 0.08 & 0.30 & 0.27\\
log([O\,{\sc i}]~$\lambda$6300/H$\alpha$) & 0.21 & 0.10 & 0.41 & 0.36\\
 \hline
\end{tabular}
\end{center}
\medskip
The mean dispersion $\mu_{\sigma}$ for the BPT line ratios with its corresponding standard deviation $\sigma$ for the star-forming and Seyfert galaxies in the sample. They are derived from the dispersion of each line ratio in every starlight-subtracted elliptical aperture, with respect to the ratios in the nuclear aperture (see Section 3).
\end{table}

\begin{figure*}
\centering
\begin{minipage}{0.8\textwidth}
	\includegraphics[keepaspectratio=true, angle=0, scale=0.8]{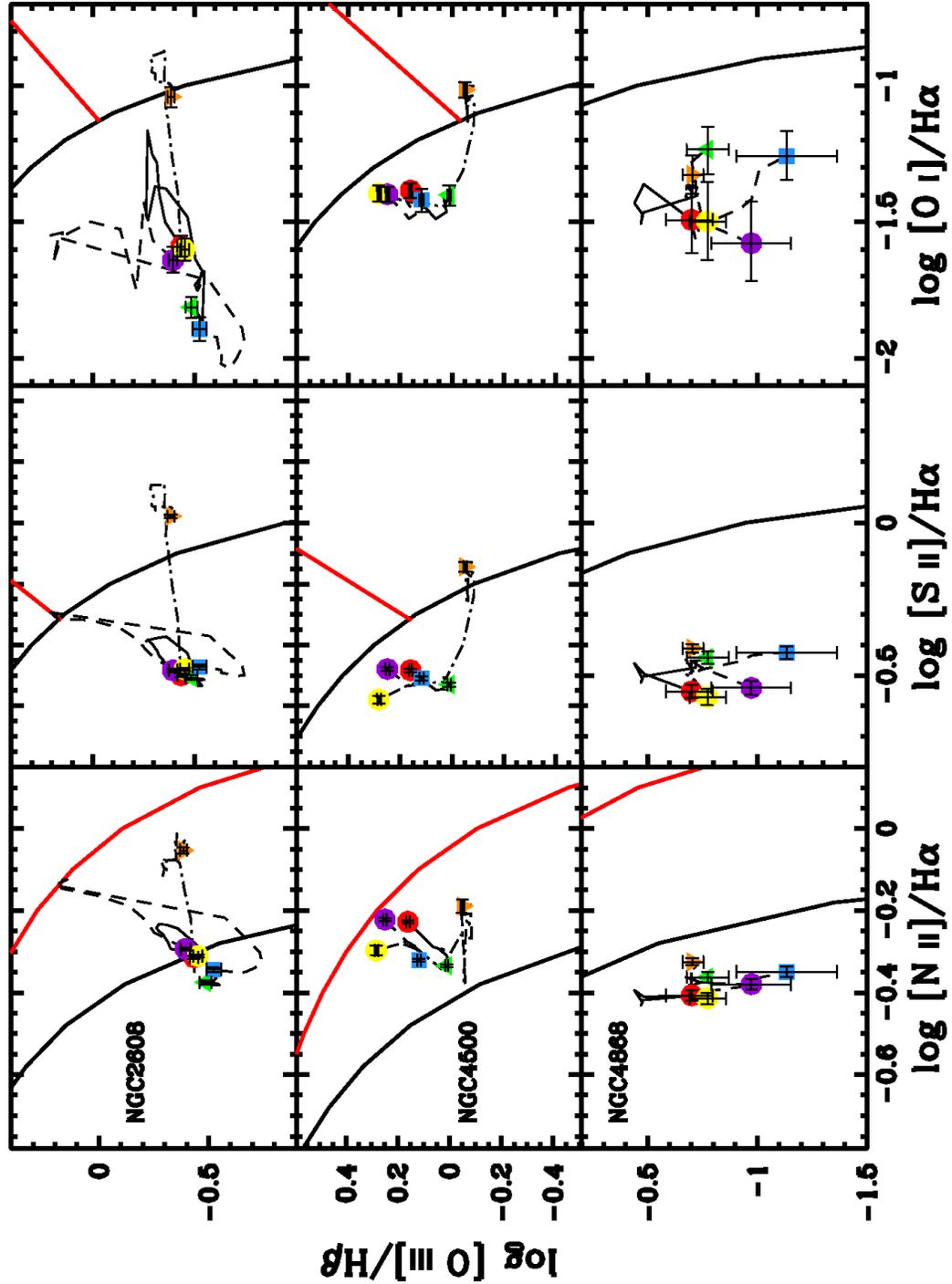}
\end{minipage}

\caption{\textbf{a} The three line-ratio diagnostic diagrams for the star-forming galaxies NGC~2608, NGC~4500, and NGC~4868, presenting a small decrease in their line-ratios with increasing aperture. The solid line represents the simulated elliptical apertures, starting from the smallest one around the nuclear region (red circle) up to the largest covering the entire galaxy (green triangle). The dashed line shows the ratios without starlight subtraction, with the purple circle indicating the spectrum from the smallest aperture and the blue square the largest one. The ratios based on the long-slit apertures is shown with the dash-dot line with the yellow circle representing the central aperture, often coincident with the central elliptical aperture, and the upside-down orange triangle the aperture along the full length of the major axis of the galaxy. On some occasions the starting points of the two methods slightly differed, because the central elliptical aperture was often 5 pixels wide and hence encloses more light than the central long-slit aperture, being always 3 pixels wide.}

\label{BPT1}
\end{figure*}

\begin{figure*}
\setcounter{figure}{1}
\centering
\begin{minipage}{0.8\textwidth}
	\includegraphics[keepaspectratio=true, angle=0, scale=0.8]{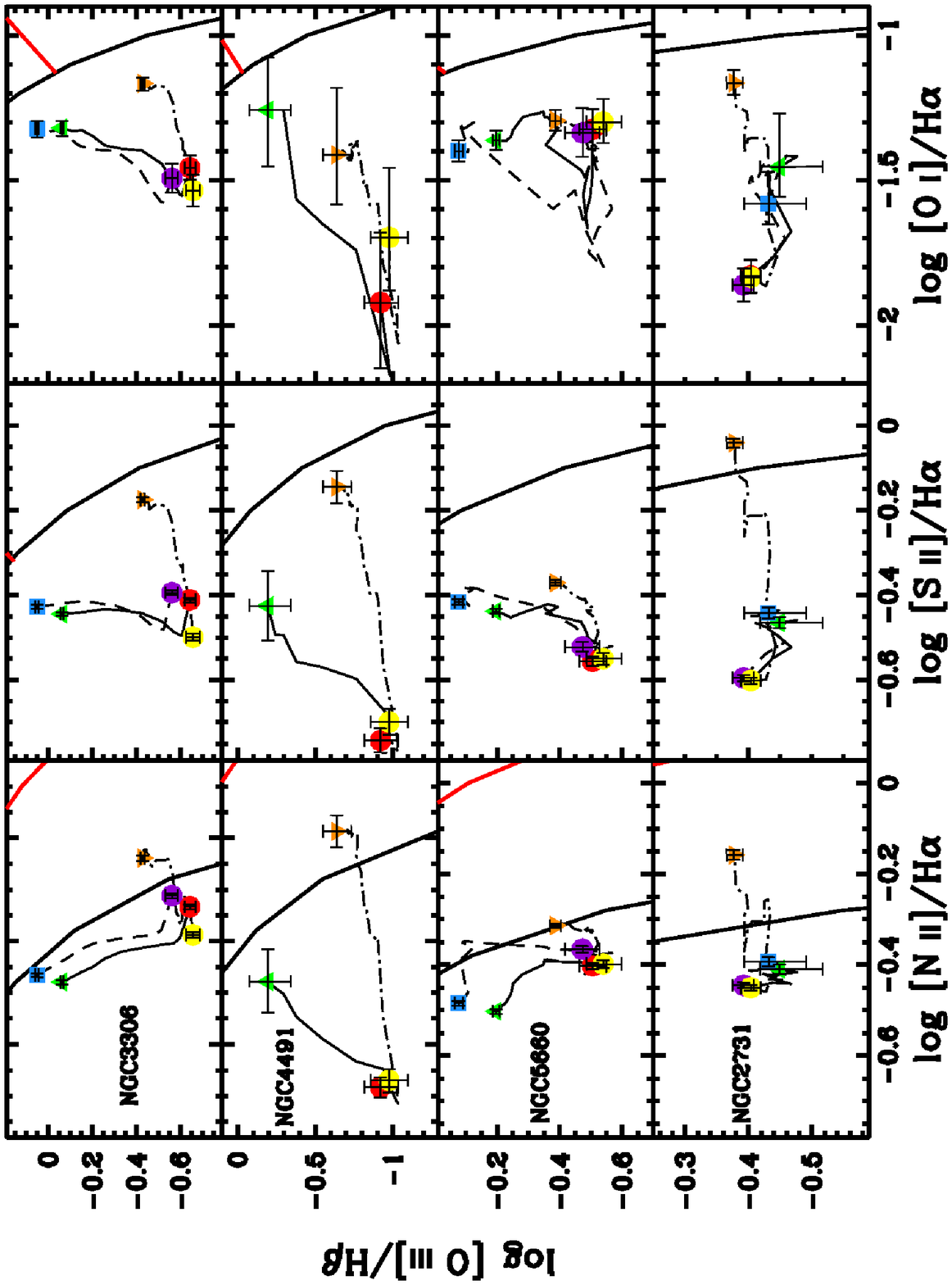}
\end{minipage}

\caption{\textbf{b} The BPT diagrams of star-forming galaxies NGC~3306, NGC~4491, NGC~5660 and NGC~2731, with the three first presenting an upward trend on the [O\,{\sc iii}]~$\lambda$5007/H$\beta$ axis with increasing aperture.}

\label{BPT2}
\end{figure*}

\begin{figure*}
\setcounter{figure}{1}
\centering
\begin{minipage}{0.8\textwidth}
	\includegraphics[keepaspectratio=true, angle=0, scale=0.8]{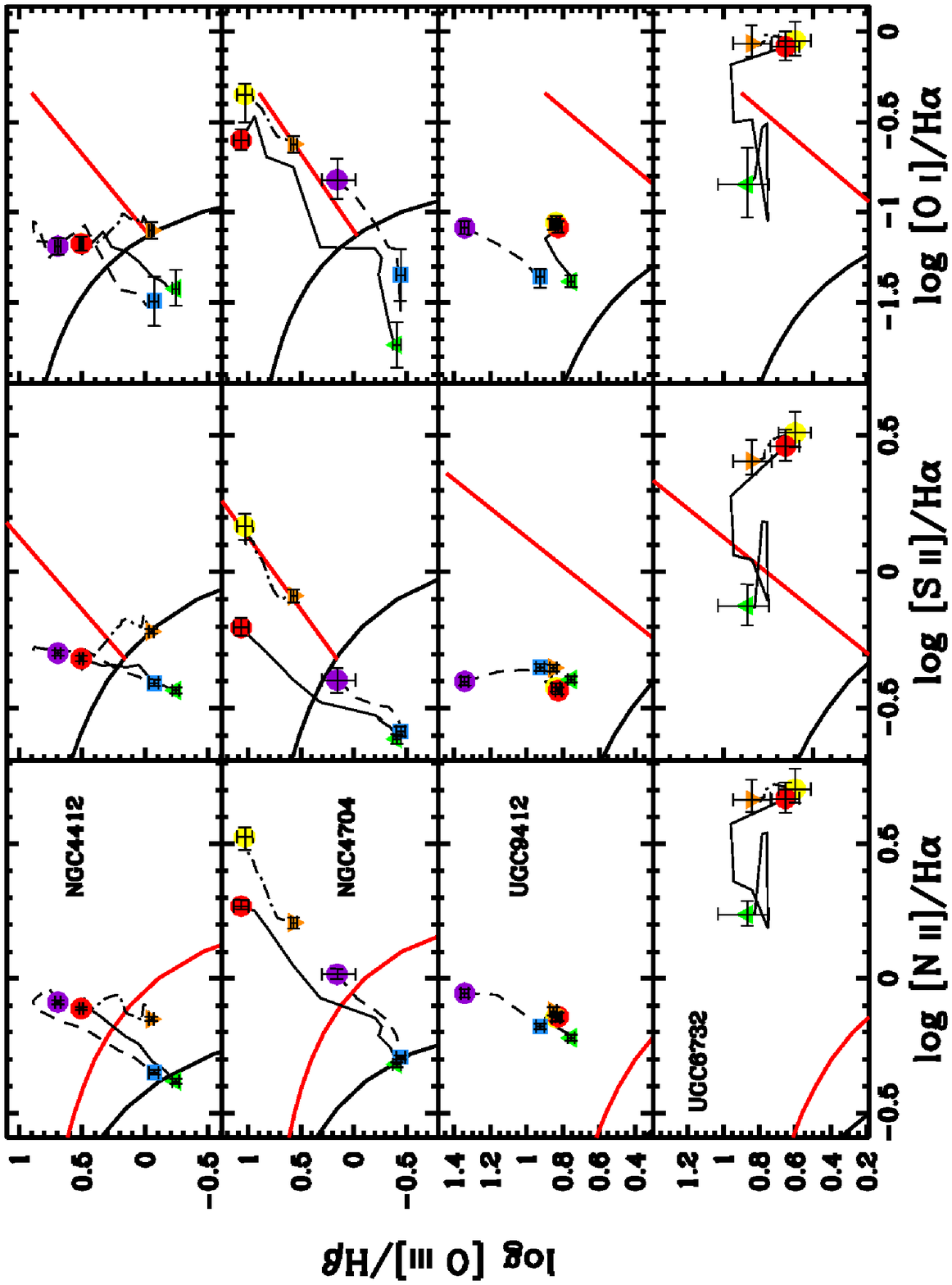}
\end{minipage}

\caption{\textbf{c} The BPT diagrams of the three spiral-Seyfert galaxies NGC~4412, NGC~4704, UGC~9412 and LINER UGC~6732. The line-ratio decrement with increasing aperture is clearly visible in the case of the Seyfert galaxies.}

\label{BPT3}
\end{figure*}

\begin{figure*}
\setcounter{figure}{1}
\centering
\begin{minipage}{0.8\textwidth}
	\includegraphics[keepaspectratio=true, angle=0, scale=0.8]{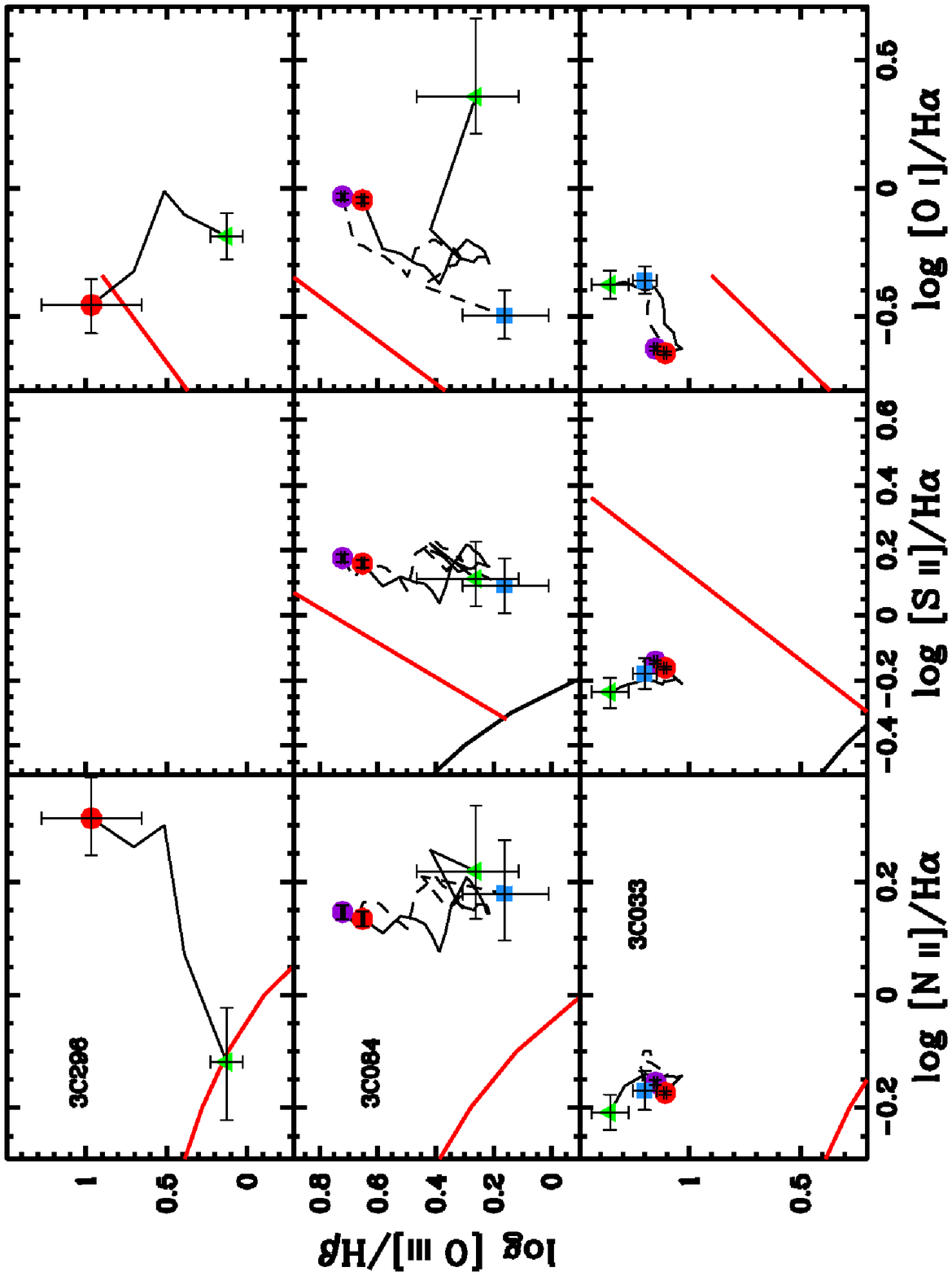}
\end{minipage}

\caption{\textbf{d} The BPT diagrams of the elliptical galaxies 3C~033, 3C~084 and 3C~296, without applying the long-slit method. Galaxy 3C~296 had no visible emission lines without starlight subtraction, while the [S\,{\sc ii}]~$\lambda\lambda$6716, 6731 doublet was not detected even after starlight subtraction; hence there is no available $[S_{II}]/H\alpha $ BPT diagnostic for this galaxy.}

\label{BPT4}
\end{figure*}

\begin{figure*}
\centering

\subfigure{\includegraphics[scale=0.66, angle=0]{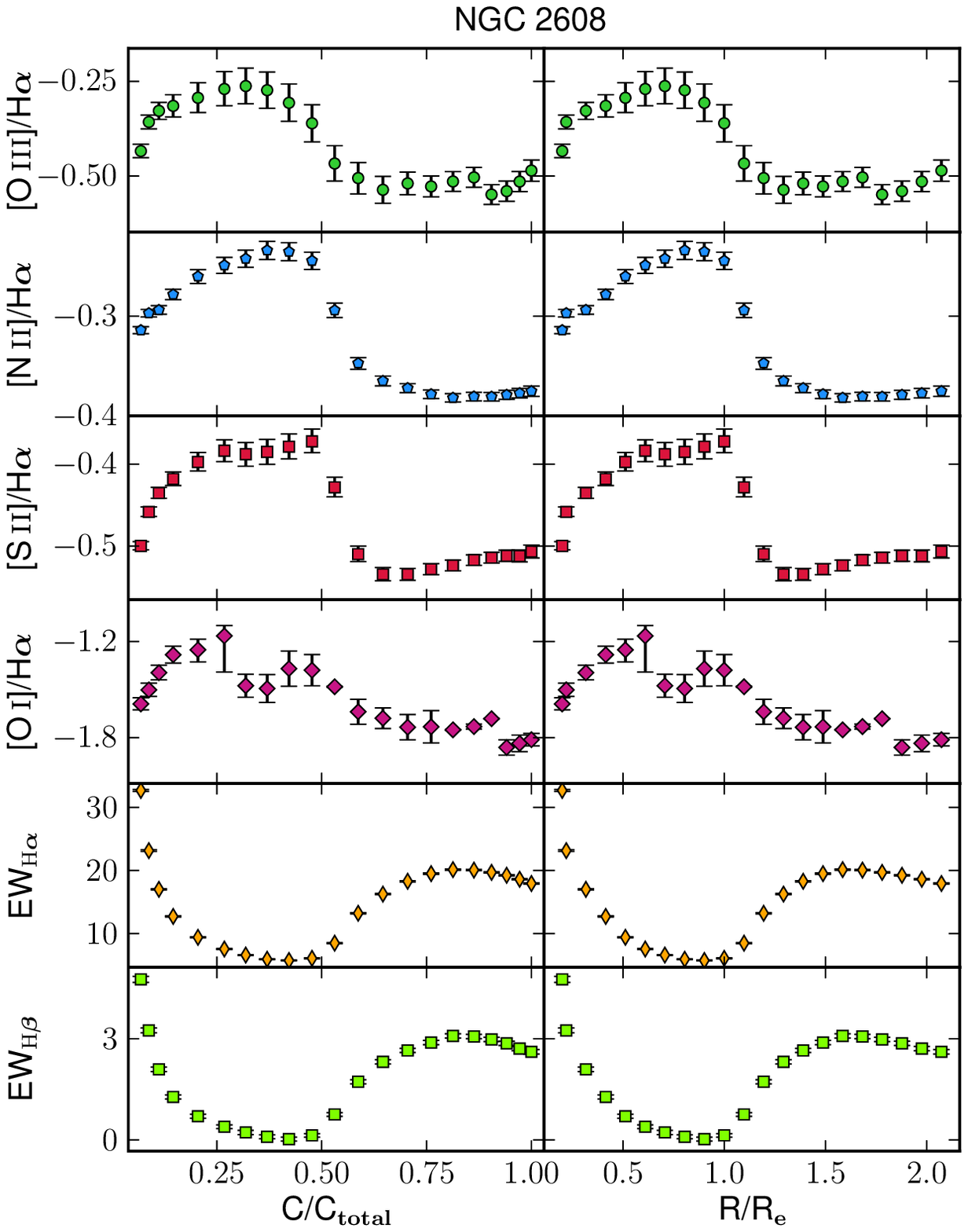}} 
\subfigure{\includegraphics[scale=0.66, angle=0]{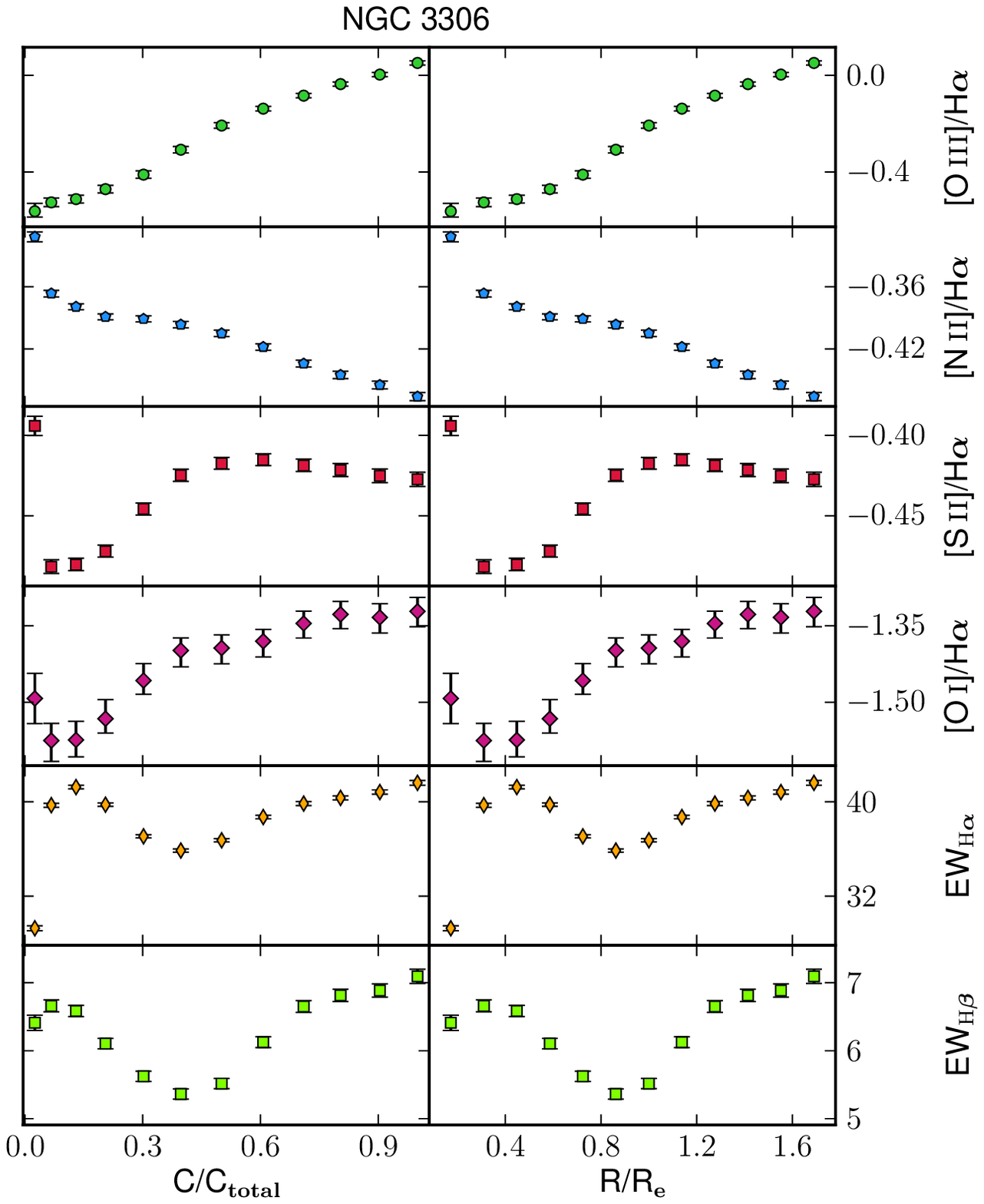}} 
\subfigure{\includegraphics[scale=0.66, angle=0]{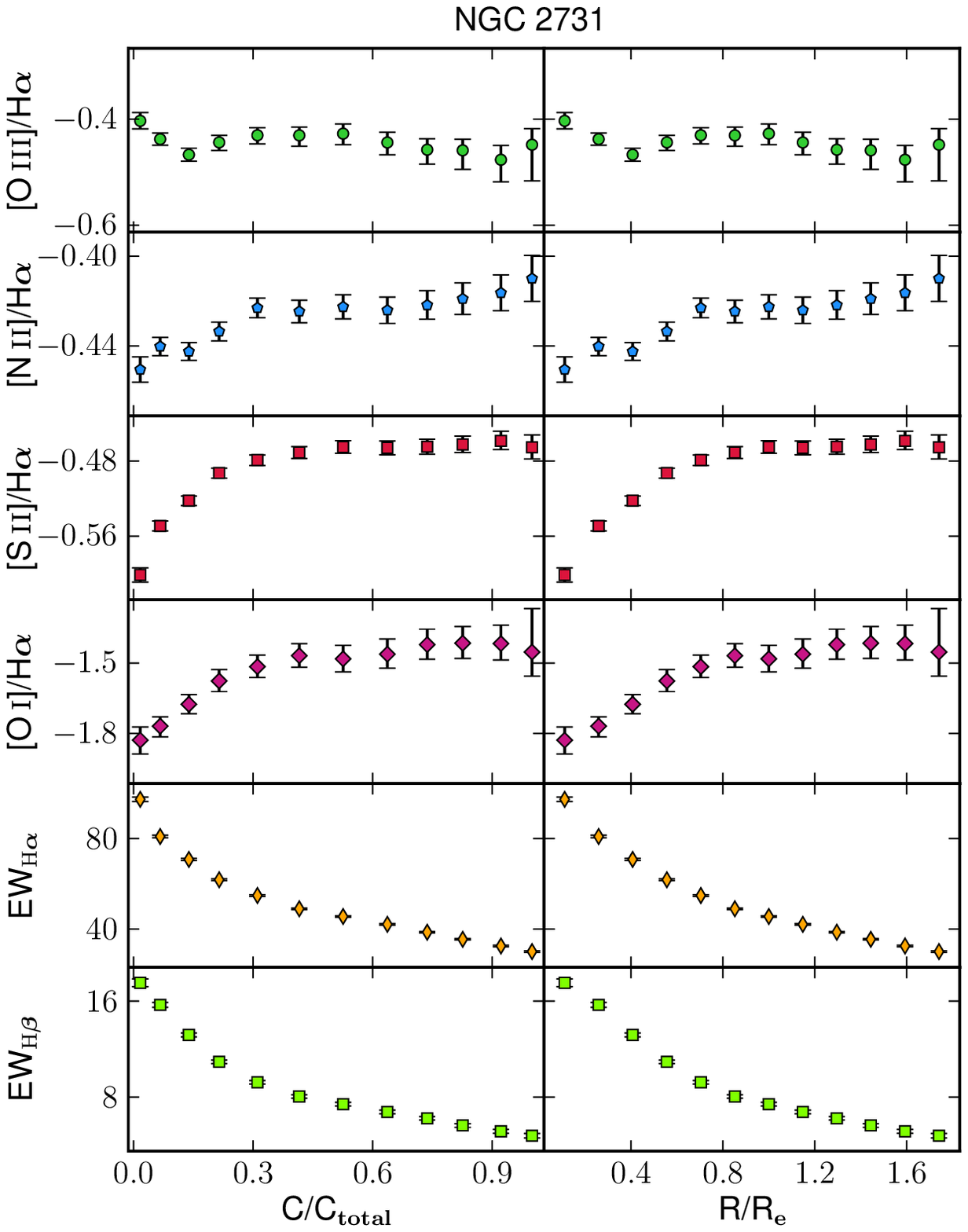}} 
\subfigure{\includegraphics[scale=0.66, angle=0]{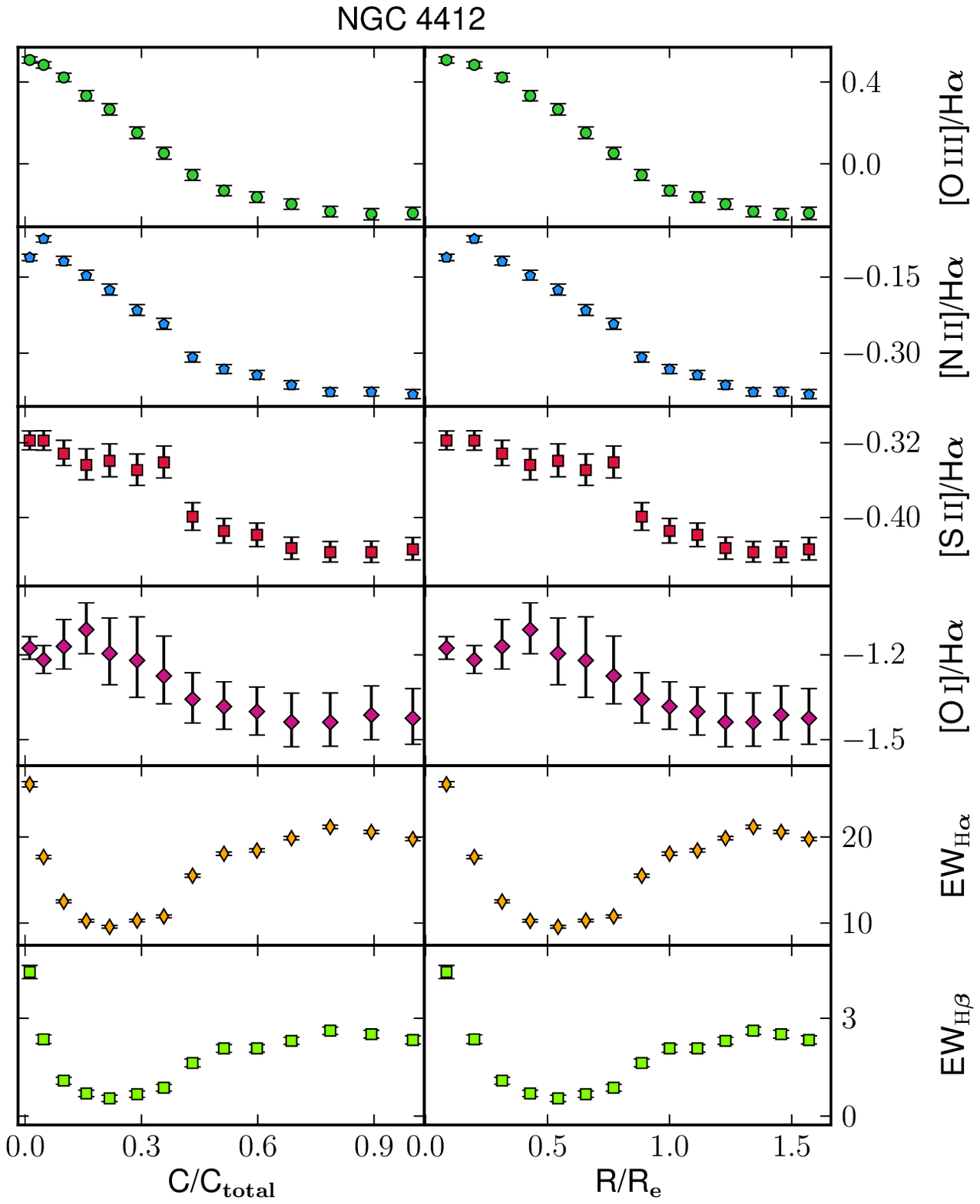}}

\caption{\textbf{a} Radial dependence of line ratios and equivalent widths measured in simulated elliptical apertures.  On the left side of each panel, the abscissa is light enclosed relative to total light of the galaxy.  The right-side abscissas are radius relative to the galaxy's half-light radius.  Line ratios are shown after starlight subtraction and given in decimal logarithms.  Equivalent widths are in {\AA}ngstrom units as observed prior to starlight subtraction. The plots of NGC~2608, NGC~2731, NGC~3306, and NGC~4412 are shown here.}

\label{6PLOTS1}
\end{figure*}

\begin{figure*}
\setcounter{figure}{2}
\centering

\subfigure{\includegraphics[scale=0.66, angle=0]{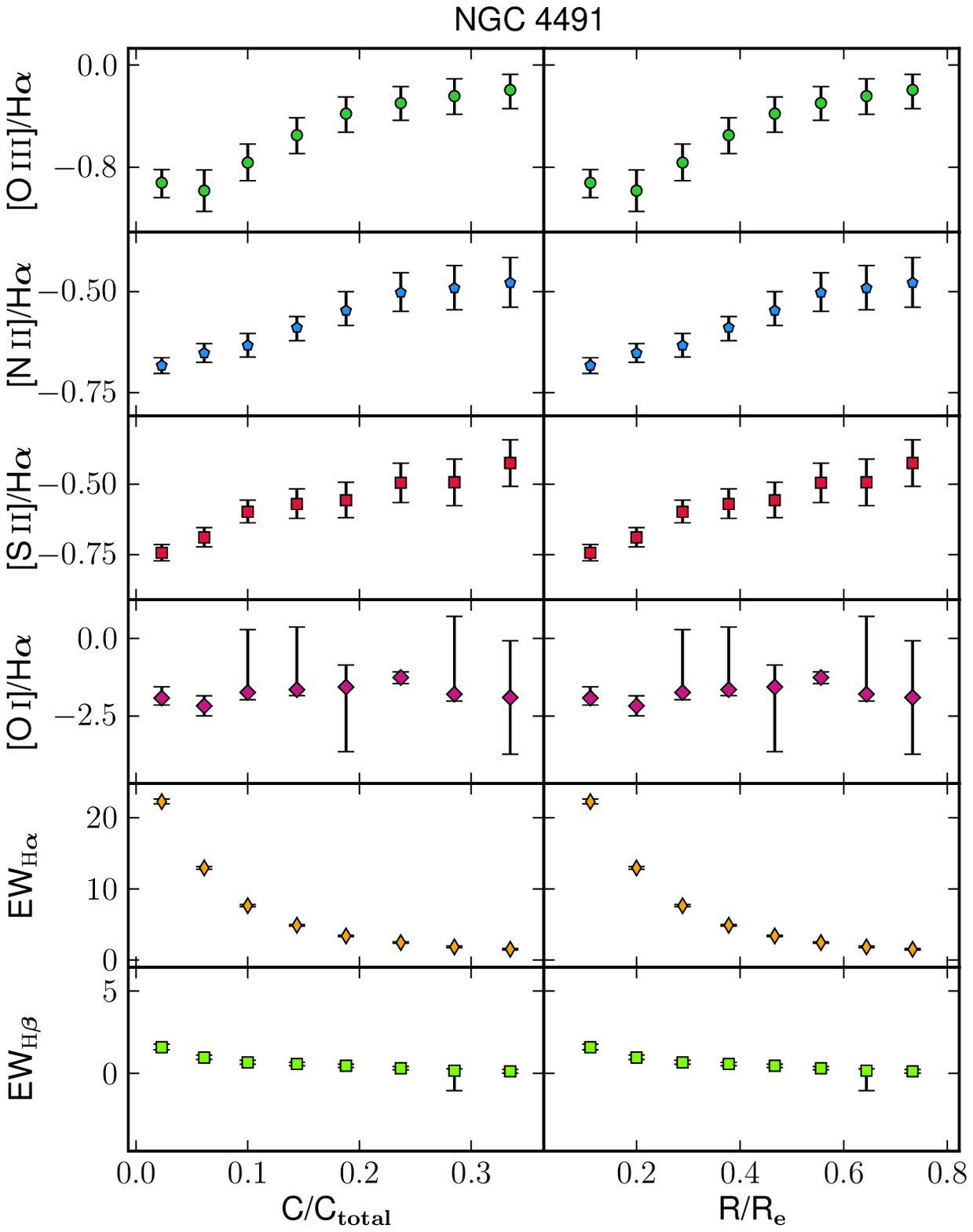}} 
\subfigure{\includegraphics[scale=0.66, angle=0]{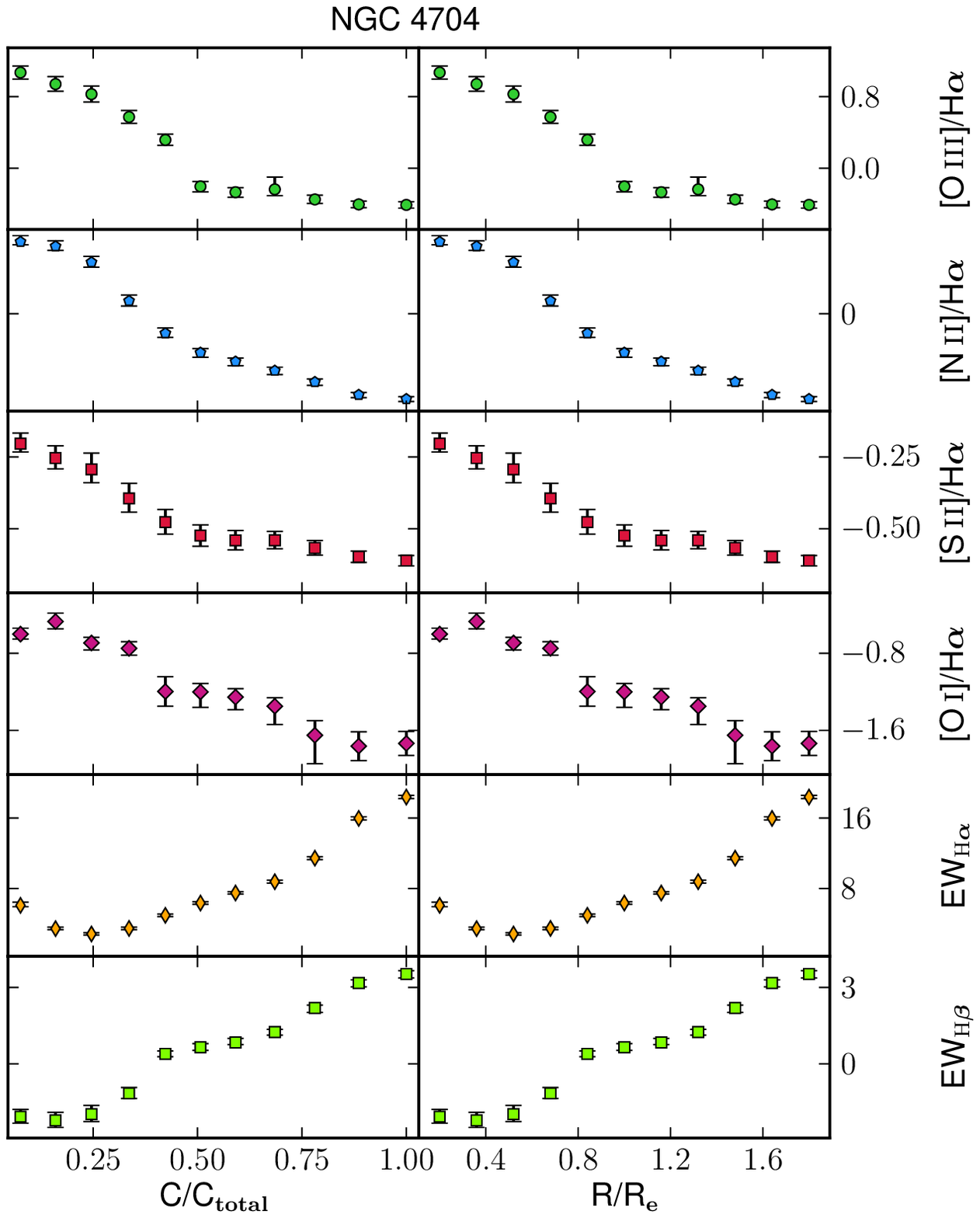}} 
\subfigure{\includegraphics[scale=0.66, angle=0]{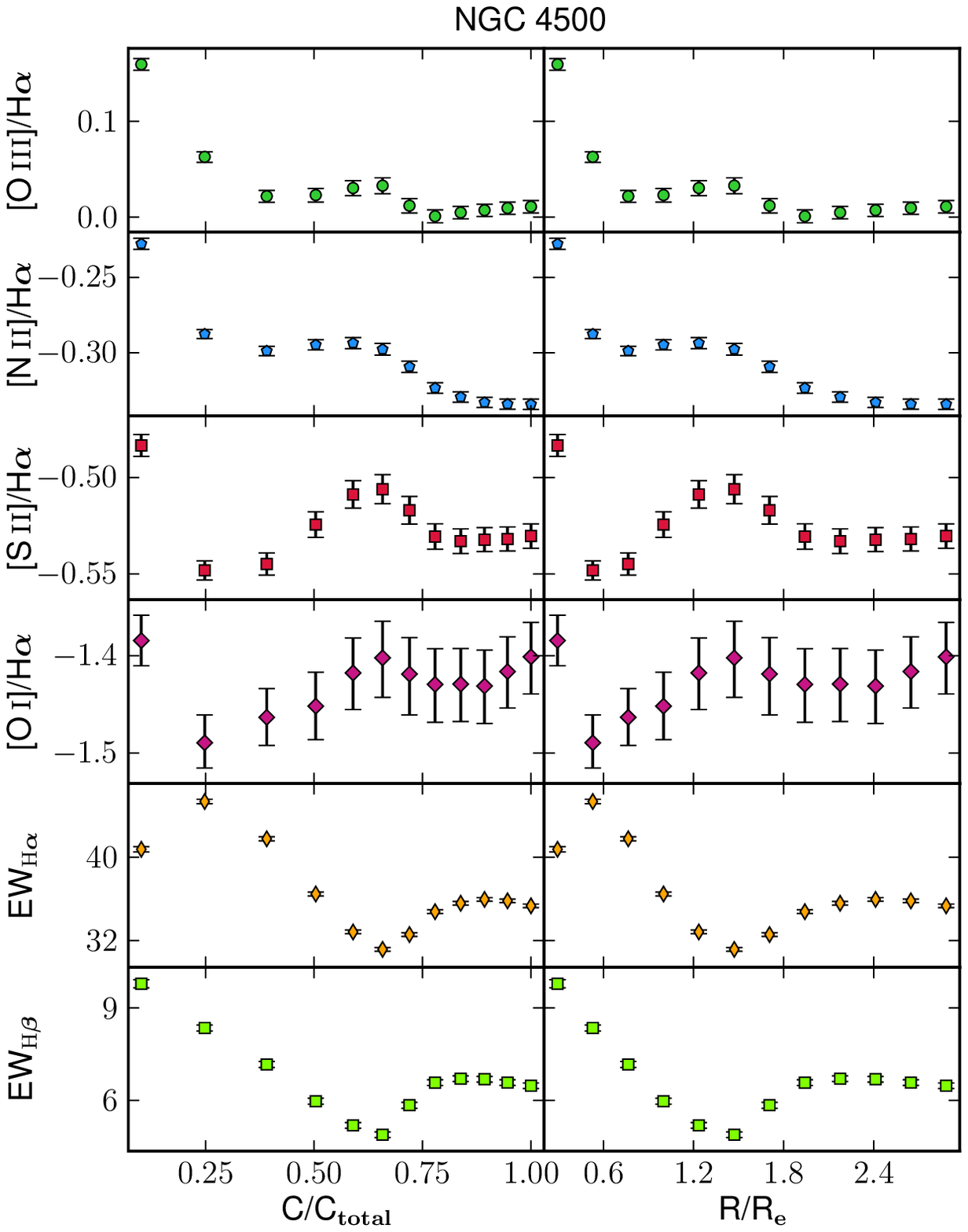}} 
\subfigure{\includegraphics[scale=0.66, angle=0]{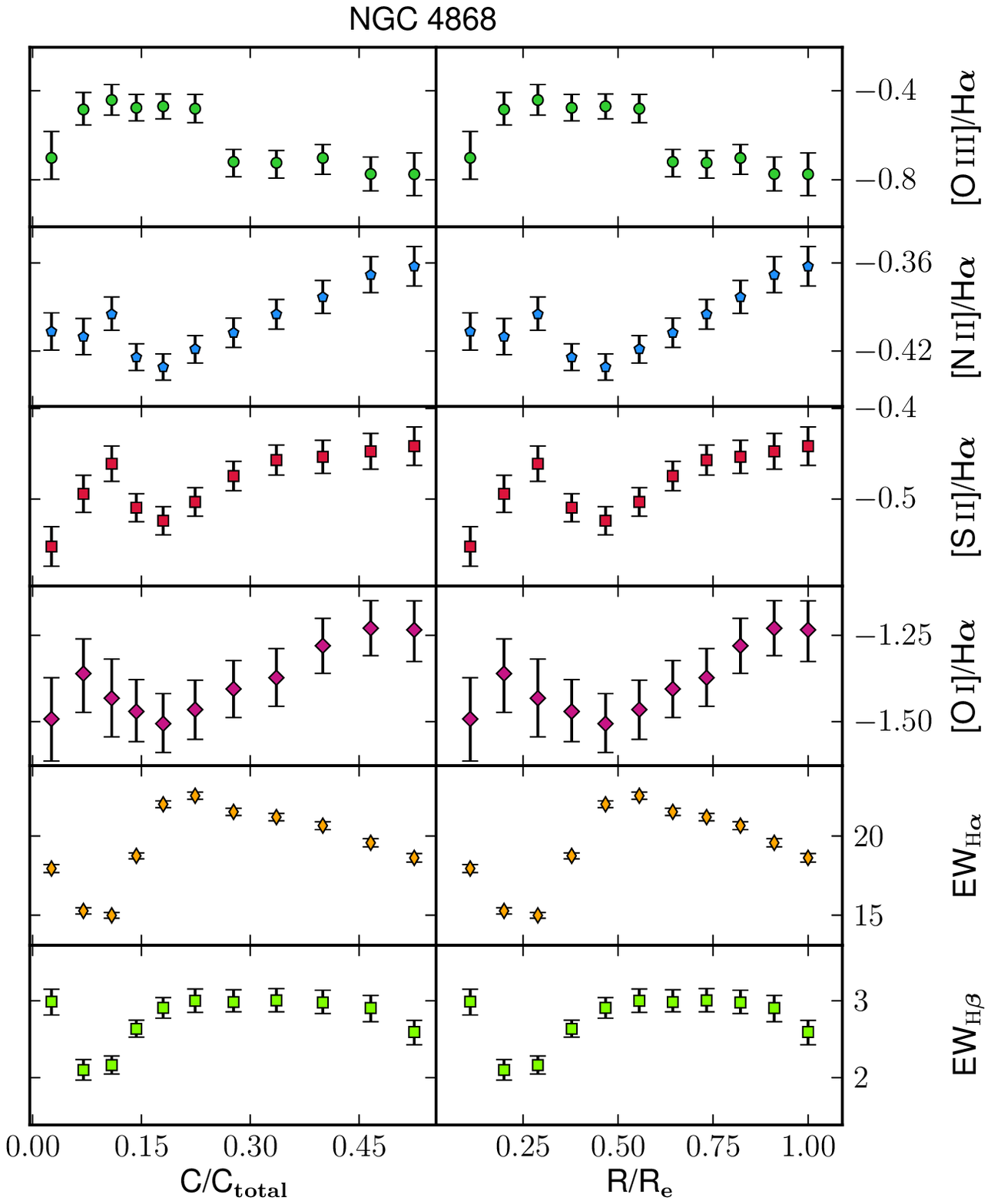}} 

\caption{\textbf{b} The plots of NGC~4491, NGC~4500, NGC~4704, and NGC~4868.}
\label{6PLOTS2}
\end{figure*}

\begin{figure*}
\setcounter{figure}{2}
\centering

\begin{tabular}{cc}
\subfigure{\includegraphics[scale=0.66, angle=0]{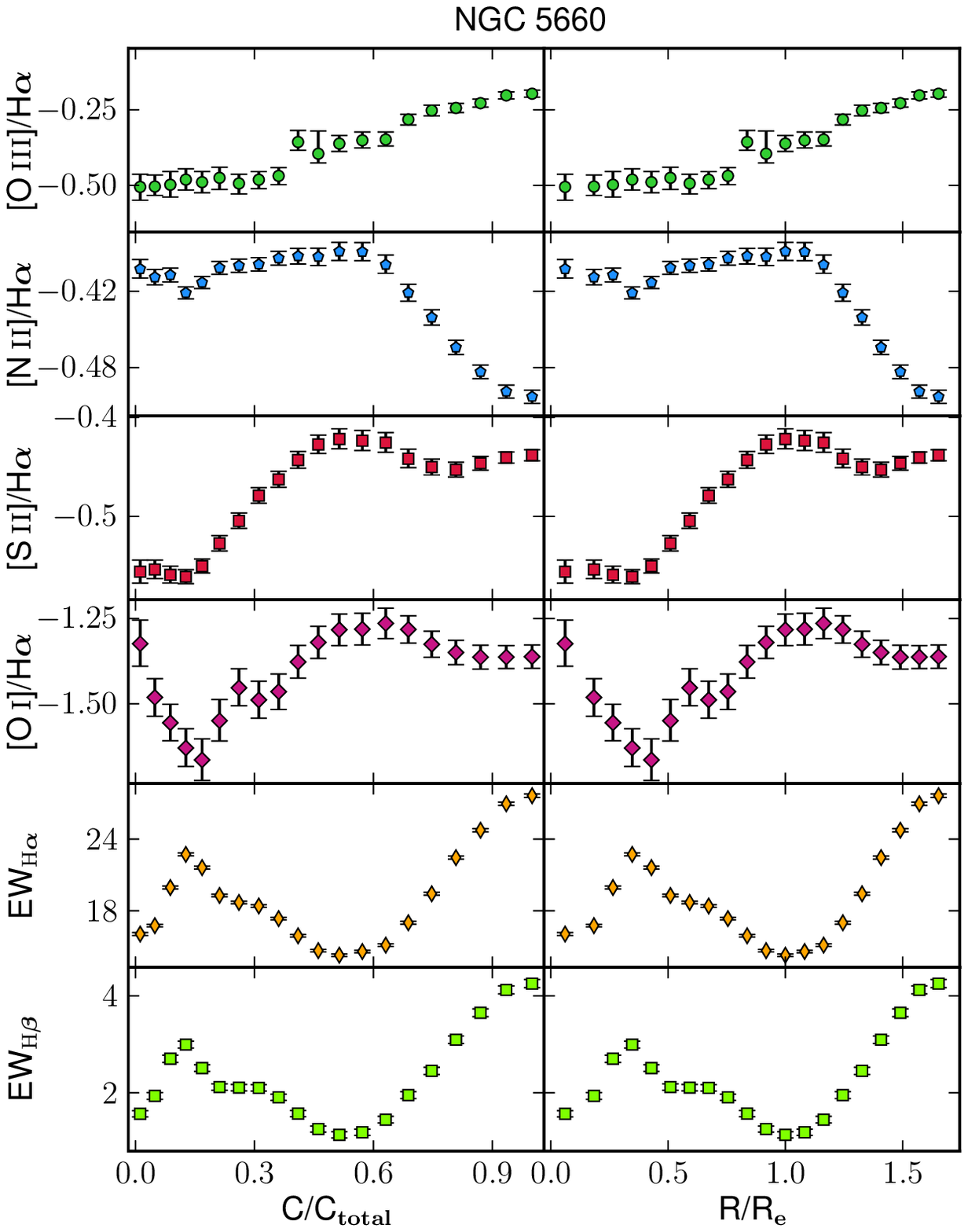}} 
&\subfigure{\includegraphics[scale=0.66, angle=0]{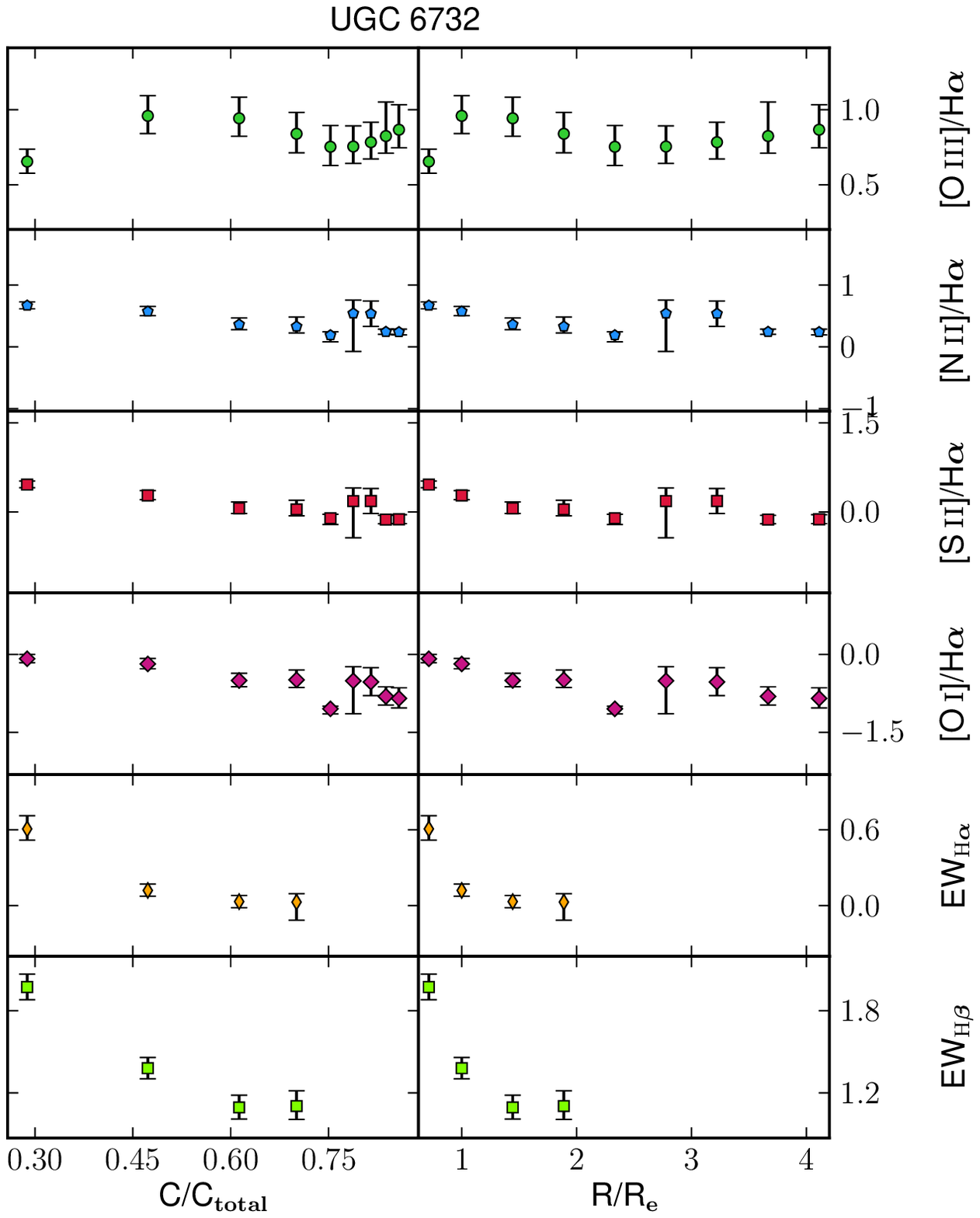}} \\
\subfigure{\includegraphics[scale=0.66, angle=0]{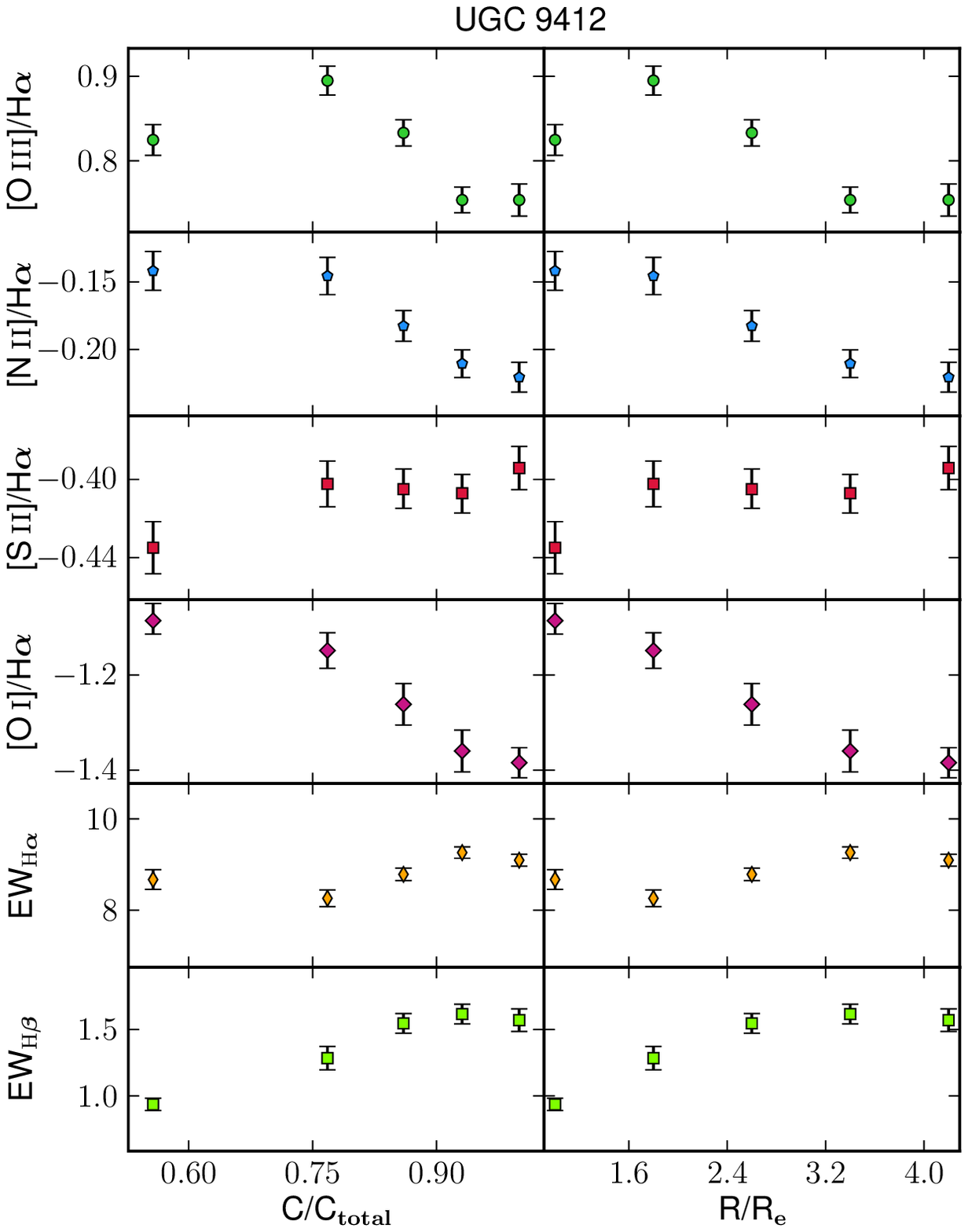}} \\
\end{tabular}

\caption{\textbf{c} The plots of NGC~5660, UGC~6732 and UGC~9412. In the case of UGC~6732 the equivalent widths of H$\alpha$ and H$\beta$ are plotted up to the point that the lines were visible.}
\label{6PLOTS3}
\end{figure*}

\begin{figure*}
\setcounter{figure}{2}
\centering

\begin{tabular}{cc}
\subfigure{\includegraphics[scale=0.66, angle=0]{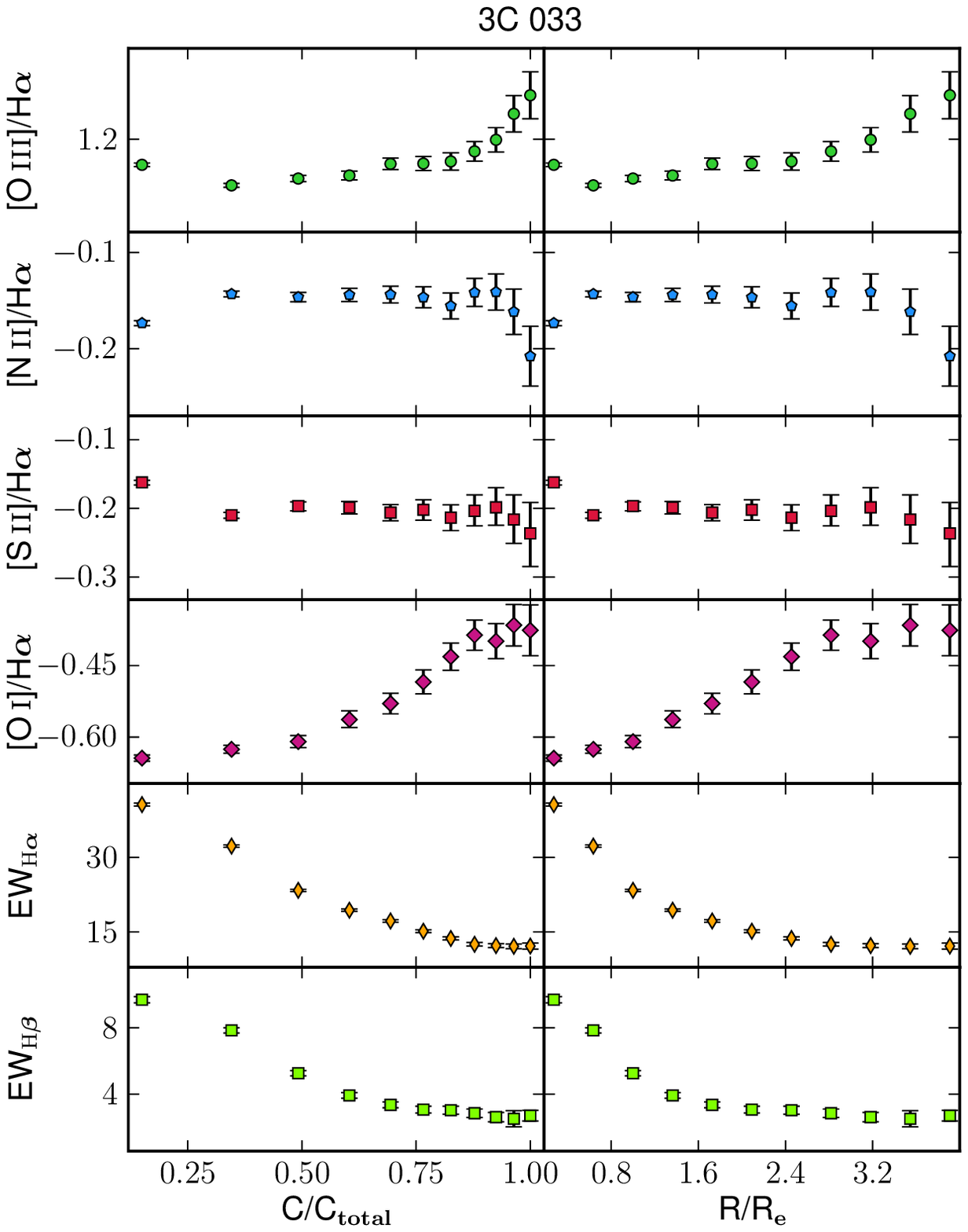}} 
&\subfigure{\includegraphics[scale=0.66, angle=0]{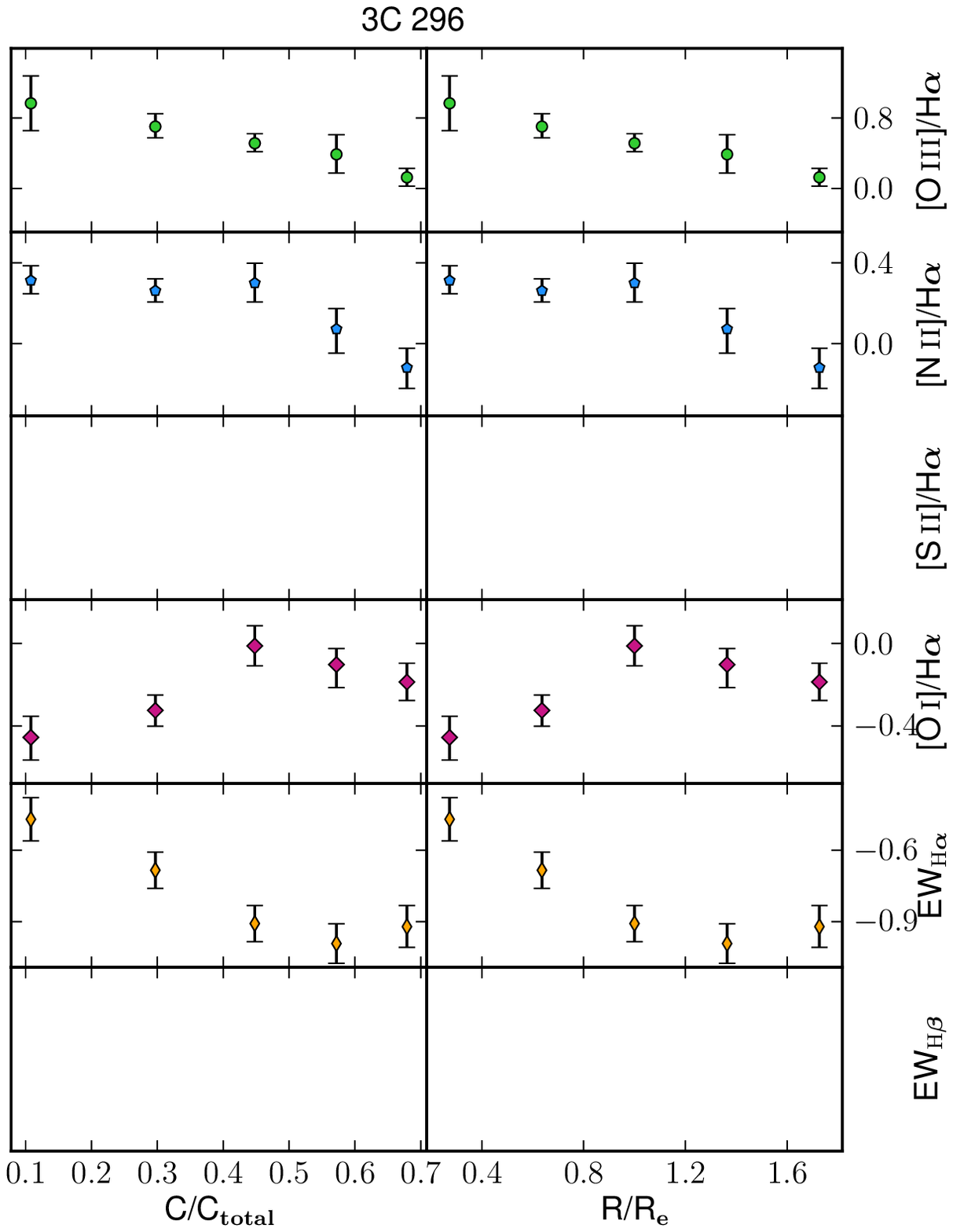}} \\
\subfigure{\includegraphics[scale=0.66, angle=0]{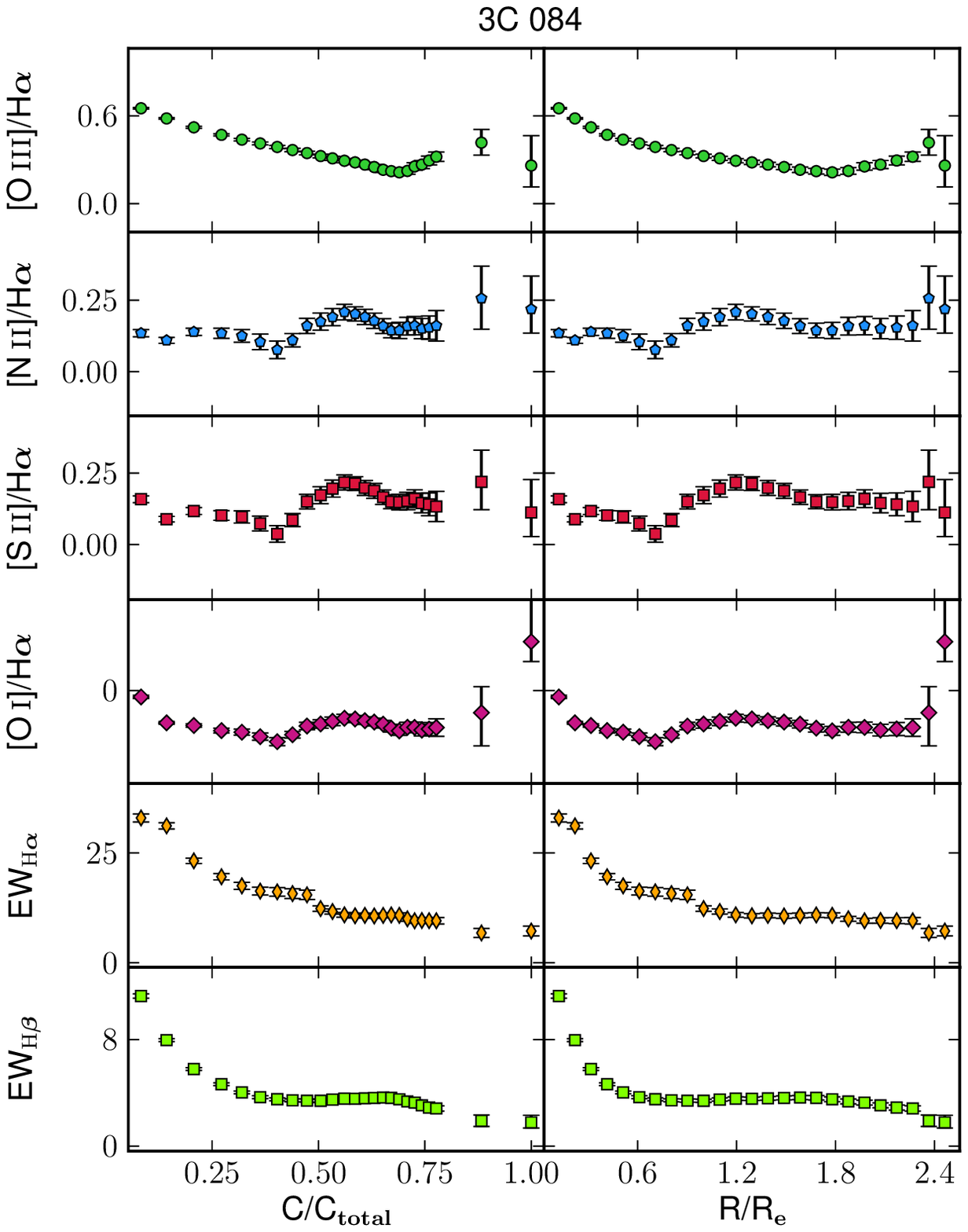}} \\
\end{tabular}

\caption{\textbf{d} The plots of the elliptical galaxies 3C~033, 3C~084 and 3C~296. No [S\,{\sc ii}]~$\lambda\lambda$6716, 6731 emission lines or H$\beta$ equivalent widths were measured for 3C296.}
\label{6PLOTS4}

\end{figure*}

\section{DISCUSSION}

\subsection{GENERAL TRENDS}

As slit length or aperture size gradually increases, there is a corresponding increase in the starlight and extra-nuclear line emission in the spectra. We expected to see all emission line intensities decrease as the proportion of starlight in the extraction aperture increased, diluting the nuclear emission (e.g., \citealp{Moran02}). In the case of galaxies hosting AGN, as we move farther away from the central regions we expected the intensities of higher ionization lines to drop faster than those of Balmer lines, as the former have usually a higher ionization potential than hydrogen. On the other hand, young or low-metallicity H {\sc ii} regions found further away from the nucleus of late-type galaxies are expected to produce high excitation emission lines. The net result will depend on both the age and the metallicity of the stellar populations on the H {\sc ii} regions. In the first case the intensities of the [O\,{\sc iii}]~$\lambda$5007 and [N\,{\sc ii}]~$\lambda$6583 emission lines should decrease faster compared to the Balmer lines. Consequently, in terms of activity classification, the respective emission-line ratios in an e.g., [O\,{\sc iii}]~$\lambda$5007/H$\beta$ - [N\,{\sc ii}]~$\lambda$6583/H$\alpha$ BPT diagram should gradually decrease when moving from the center to the edge of galaxies. This will produce a classification trend towards the H\,{\sc ii} region. This is exactly what we observe, both with and without starlight subtraction, in the case of the spiral-Seyfert galaxies NGC~4412, NGC~4704, and less strongly for UGC~9412, in agreement with the results of Moran et al (2002). The same trend in the BPT diagrams is also observed in the case of the elliptical-Seyfert galaxies 3C~084 and 3C~296. Because the ionization potential of [O\,{\sc iii}]~$\lambda$5007 is twice that of the other emission lines of interest, it is expected to drop more abruptly following the reduction of high-energy photon density as we move outwards in the galaxy. Indeed Figures \ref{BPT1} show the larger variations along the [O\,{\sc iii}]~$\lambda$5007/H$\beta$ axis as opposed to the horizontal axis of the diagrams.


\subsection{TRENDS IN LATE-TYPE GALAXIES}

For the spiral galaxies, we see two different trends. Some galaxies (NGC~4500, NGC~4868) show lower [O\,{\sc iii}]~$\lambda$5007/H$\beta$ values with increasing aperture, in line with our expectations. On the other hand, four galaxies (NGC~3306, NGC~4491, NGC~5660, and UGC~6732) show ratios moving towards the upper, higher excitation regions of the diagnostic diagrams. This behavior is apparent in all types of analysis (long-slit, elliptical aperture with/without starlight subtraction). Of these galaxies NGC~3306, NGC~4491, and NGC~5660 are classified as star-forming, while UGC~6732 is classified as LINER based on their long-slit spectra of their nuclear (3.5$\arcsec$ or 5.85$\arcsec$ x 3$\arcsec$) regions. Star formation does not occur exclusively in the nuclei of galaxies but is distributed throughout the galactic disk \citep{Kenn89}. Star-formation activity of different ages, occurring in various areas throughout the galaxy, will produce ionizing continua of different shapes with more recent starbursts giving the most energetic photons. As seen in the two-dimensional spectra and the SDSS $g$-band and GALEX near-UV images of galaxies NGC~3306 and NGC~5506 (Figure \ref{NGC5660-O-UV}), several star-forming clumps are present on their spiral arms. Our spectra of these knots shows that their [O\,{\sc iii}]~$\lambda$5007-line intensity is greater than that of the central star-forming region, indicating a harder ionizing continuum and hence younger stellar populations. A similar effect could arise from metallicity gradient observed in galaxies. These effects can explain the increasing excitation of NGC~3306 and NGC~5660 (Figure \ref{BPT2}b). The result highlights how morphology and galaxy structure can affect the analysis and the activity classification even of pure star-forming galaxies even though the line ratios remain within the star-forming range at all radii.

Interarm regions of galaxies should enhance the overall starlight component of the spectrum because interstellar gas is mostly absent from those regions. Even if there is any small amount of gas, the evolved stars present there are not capable of ionizing it. With respect to the integrated spectrum, this means that the spectral absorption features will intensify while the emission lines will remain almost constant. When applying starlight subtraction, the additional stellar populations encountered in the interarm regions can introduce complications to the fit, leading to an overestimation (or underestimation) of the starlight continuum. In turn, this would lead to a larger (or smaller) estimation of the Balmer absorption line equivalent widths, and the resulting subtraction of the stellar continuum would have an impact on the flux of Balmer emission lines. Lower values of the H$\beta$ emission line results in higher placement of the galaxy in all BPT diagrams. Furthermore, when constructing the elliptical apertures, considering that the elliptical annulus size increases as we move towards the outer regions of a galaxy while the rectangular regions always maintain a fixed size, the resulting weight used to multiply the rectangular region's spectrum will increase with respect to the radius. In the case of the interarm regions this will produce a spectrum of intense absorption features and increasing starlight continuum resulting in reduced apparent intensity of the H$\beta$ emission line. This is applicable not only in interarm regions but also in the extra-nuclear regions of galaxies with low star-forming activity in general. This could be the case in NGC~4491, where the quality of the elliptical simulated spectra dropped dramatically as we expanded the apertures towards the galaxy edges, and we were able to measure the emission line ratios out to approximately half of the surface area. This effect could be the source of the apparent (but not significant) upward trend for NGC~4491 on the diagnostic diagrams. Correspondingly, this could also apply for UGC~6732 where the initial (nuclear) and the final (simulated integrated) spectra have similar [O\,{\sc iii}]~$\lambda$5007/H$\beta$ ratio within their uncertainties.

Regarding the behavior of galaxies along BPT horizontal axis, excluding NGC4491 and NGC4868 which give simulated elliptical aperture spectra of lower quality, the rest of the sample tends to move towards the left side on the diagnostics as expected. The lower ionization potential of [N\,{\sc ii}]~$\lambda$6583, [S\,{\sc ii}]~$\lambda\lambda$6716, 6731 and [O\,{\sc i}]~$\lambda$6300/H$\alpha$, compared to [O\,{\sc iii}]~$\lambda$5007, allows less energetic photons to ionize these species in gaseous regions farther away from the center and hence generate a less dramatic reduction in their line ratios, resulting in smaller shifts along the horizontal axis of all three diagnostic compared to the  [O\,{\sc iii}]~$\lambda$5007/H$\beta$ variations.

For three galaxies in the sample (NGC~2608, NGC~4412, and NGC~4704), we observe an actual change in their activity classification based on the elliptical-aperture simulation method in at least one of the three diagnostic diagrams, while three more galaxies (NGC~2731, NGC~4491 and NGC~4500) have a change in their classification based on the varying standard-long-slit method. NGC~4412 and NGC~4704, classified as Seyfert based on their nuclear spectra, fall into the star-forming region of all three diagrams. (If we omit starlight subtraction they fall into the TO region in the $[N_{II}]/H\alpha$ diagnostic). Finally, NGC~2608 along with NGC~4412 and NGC~4704 undergo a change in classification towards the TO region in the [N\,{\sc ii}]~$\lambda$6583/H$\alpha$ diagnostic at an aperture covering approximately half of their surface. This is demonstrating an intermediate stage of the aperture effect that can be encountered when observing with a fiber, where a portion of the galaxy including the nucleus is observed rather than just the central region or the total surface.

\begin{figure*}
\begin{center}
\begin{tabular}{cc}
\includegraphics[keepaspectratio=true, scale=.15]{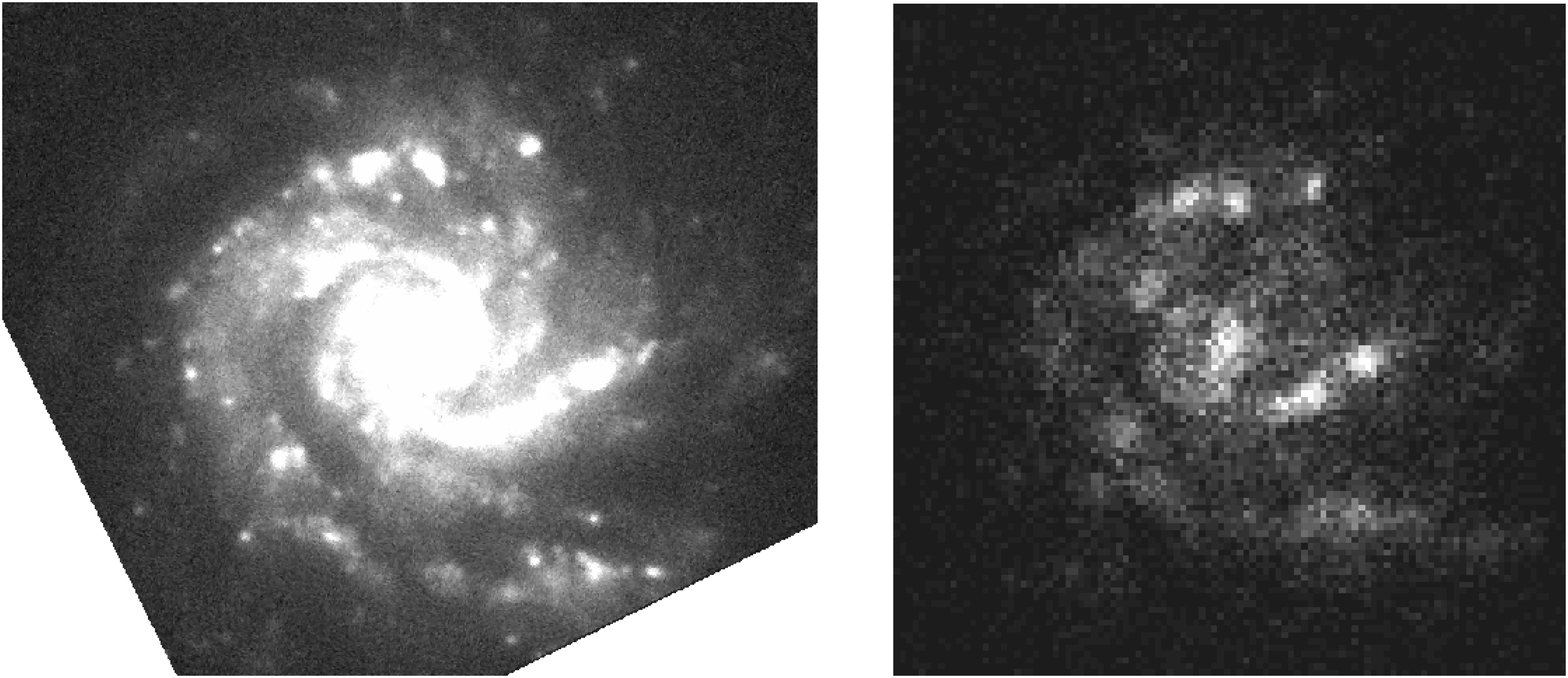} &
\includegraphics[keepaspectratio=true, scale=.165]{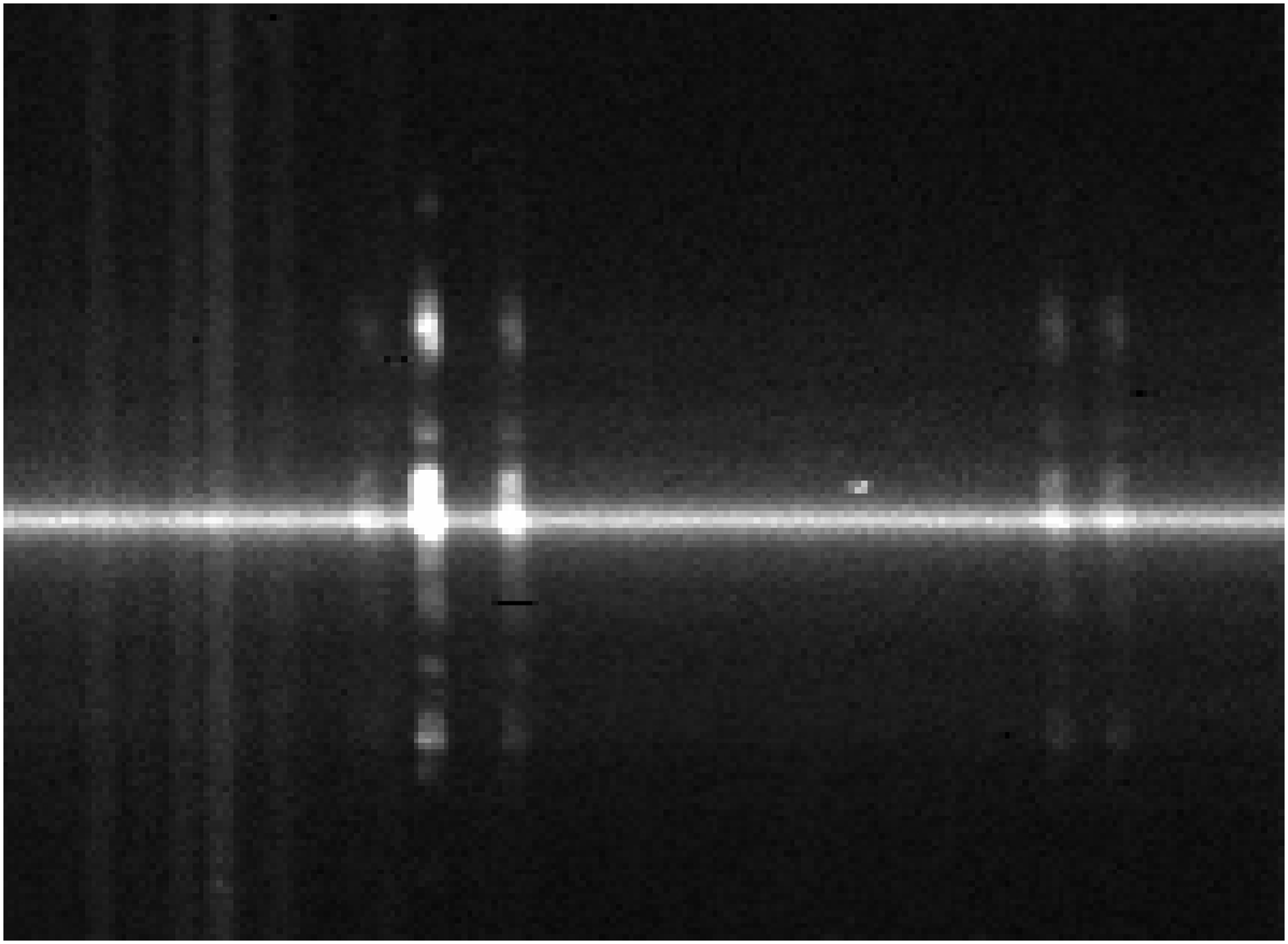} \\
\includegraphics[keepaspectratio=true, scale=.15]{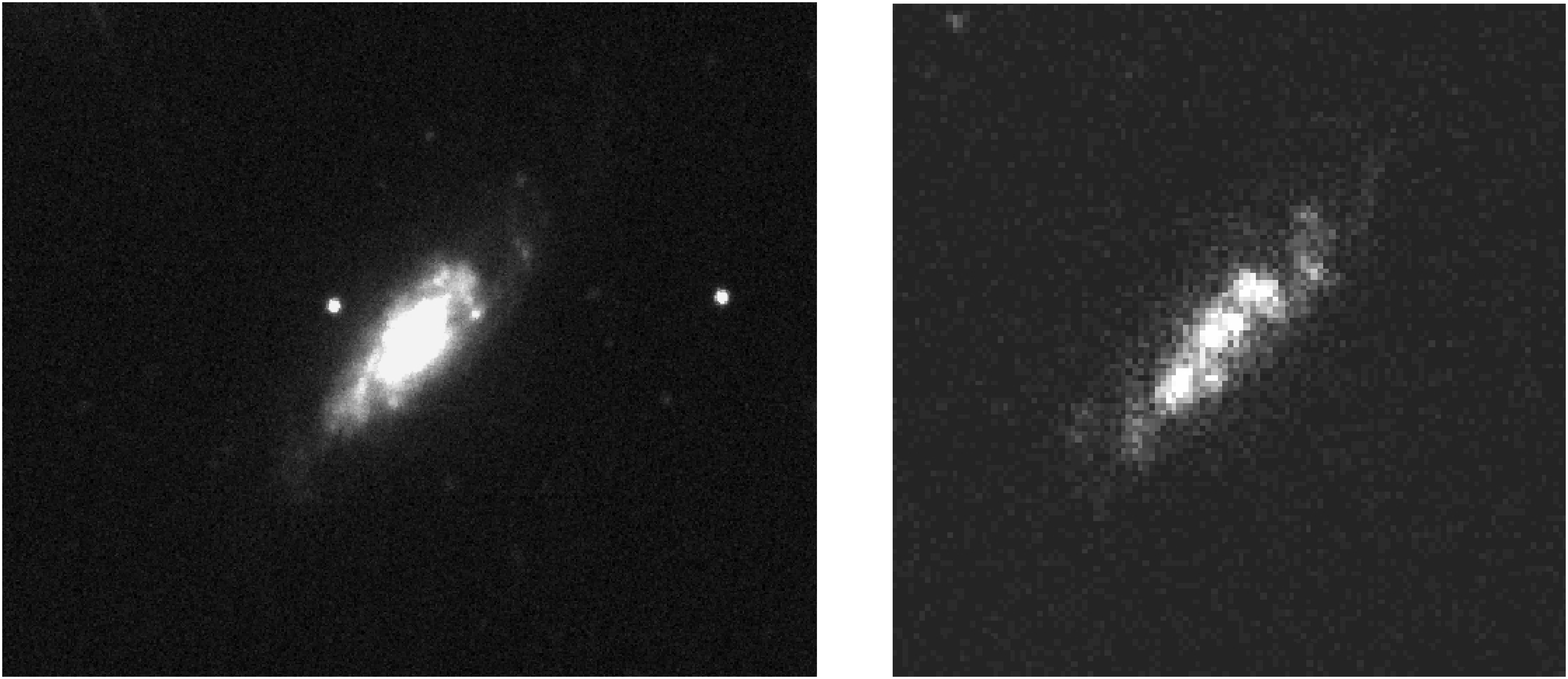} &
\includegraphics[keepaspectratio=true, scale=.23]{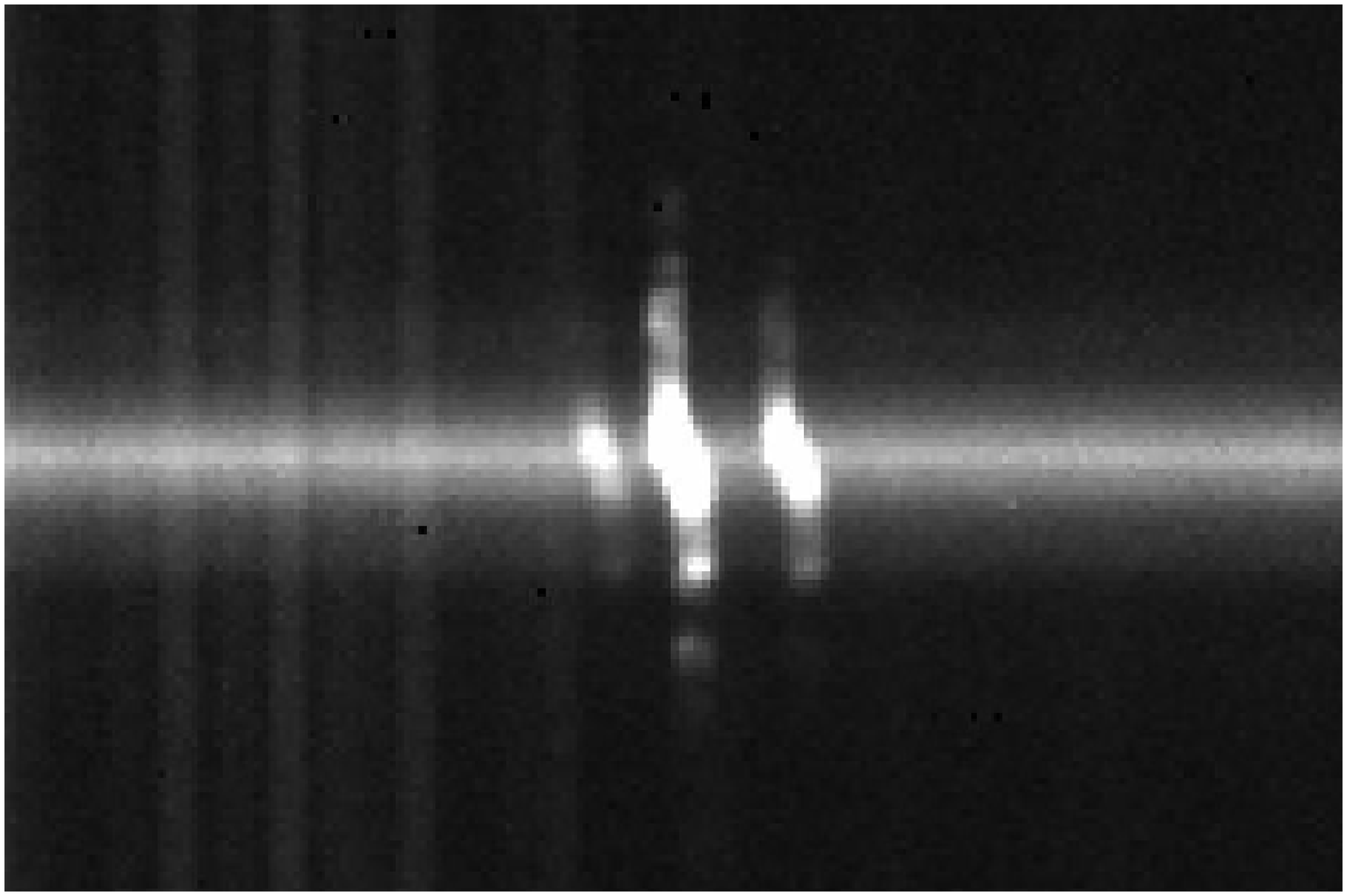} \\
\label{NGC5660-O-UV}
\end{tabular}

\caption{Upper panel: SDSS's G-band (left), GALEX UV (middle), and two-dimensional spectrum around the H$\alpha$ region for galaxy NGC~5560. The optical and UV images show star-formation clumps, while in the two-dimensional spectrum we see emission from regions away from the center. Lower panel: The same information for galaxy NGC~3306.}

\label{NGC5660-O-UV}
\end{center}
\end{figure*}

\subsection{TRENDS IN EARLY-TYPE GALAXIES}

Early type galaxies are nearly devoid of star-forming activity, simplifying in principle the complications found in late-type galaxies with extra-nuclear star formation. The sample of elliptical galaxies follows the same trend as spiral-Seyfert galaxies, that is towards lower excitation with increasing radius (Figure \ref{BPT4}d). Again, the upward tendency of 3C~033 on the optical diagnostic diagrams can be attributed to the uncertainties in the emission-line measurements. 3C~296 appears to have an activity classification change from Seyfert to TO and LINER in the case of [N\,{\sc ii}]/H$\alpha$ and [O\,{\sc i}]/H$\alpha$ diagnostics respectively. For this galaxy, we were not able to measure emission-lines without applying starlight subtraction, and even so the [S\,{\sc ii}]$\lambda$6716, 6731 doublet was not detected, hence no classification was acquired using the [S\,{\sc ii}]/H$\alpha$ diagnostic. The most intriguing aspect of this analysis for the elliptical subsample is that even when star-forming activity had no significant impact on the continuum subtraction, the host galaxy’s starlight could not be accurately removed. Galaxies fall lower in the BPT diagnostics as aperture increases even when starlight subtraction is applied, revealing the masking of the AGN features by the host galaxy's starlight.

\subsection{THE EFFICIENCY OF STARLIGHT SUBTRACTION}

One might naively expect optimal subtraction of galaxy starlight to place the central and the larger aperture sub-spectra at the same location on the BPT diagrams. The previous Section demonstrates that this is not the case. We attribute this to two different reasons. First, as we showed, star-forming activity occurring outside the nucleus of a galaxy will influence the emission from the interstellar medium and hence the resulting classification in the BPT diagrams. This is a problem that starlight subtraction cannot account for. The second reason is the possible errors of starlight subtraction. Uncovering the actual stellar populations of galaxies responsible for the continuum light diluting the emission lines is a challenging task even in the case of elliptical galaxies, which generally have simpler stellar populations than spiral galaxies. The stellar continuum is the result of the radiation from stars of different ages and metallicities. Due to the known age-metallicity degeneracy (\citealp{Worthey}), different combinations of stellar populations can produce similar continuum shapes. Furthermore, the observed spectral shape and characteristics (e.g., the FWHM of absorption lines) are also affected by the stellar velocity dispersion and extinction, both important parameters used when fitting the actual spectrum of a galaxy. Most importantly the limited number of stellar lines available after excluding the Balmer lines does not allow for an accurate and unique decomposition of the stellar component. In this respect methods based on fitting elliptical galaxy templates (e.g., \citealt{Ho97a}) rather than combination of single stellar population (e.g., STARLIGHT) could be more robust, but they do not ensure a reasonable fit in every case.

Determining all the parameters affecting a galaxy spectrum is not a trivial process. Even small deviations from the actual parameters will produce an inaccurate stellar continuum, meaning that the subtraction from the observed spectrum may not reveal the proper emission-line characteristics. This will lead to the observed displacement in the BPT diagrams between the central and all intermediate and larger aperture spectra.

For galaxies hosting AGN there is an additional parameter influencing the result of the stellar component subtraction, and that is the inclusion or omission of the AGN continuum in the fit. Generally, a power-low component is used to fit the AGN continuum. The proper selection of the power-law index though is not a standard task as it can differ between galaxies. Consequently, without accounting separately for the AGN continuum, it is treated  in the fit as part of the stellar continuum. This means that additional stellar populations may be used to supplement for the AGN continuum, producing a less accurate synthetic spectrum for subtraction. This has a major effect in the estimated Balmer absorption lines intensities from the starlight. In particular, the true Balmer absoprtion depth will be less than the calculated values, and the recombination line fluxes will be over-corrected.

\section{CONCLUSIONS}

Our analysis indicates that line ratios and consequently activity classification obtained from the SDSS 3$\arcsec$ fibers can be affected by aperture effects. The proportions of nuclear and disk (stellar) light that enter an SDSS fiber depend on galaxy distance. Thus, AGN features' detectability are likewise a function of distance. The result could be a different activity classification than the one obtained from observations specifically from the central regions. This is crucial, because from the AGN identification standpoint one must solely concentrate on the central regions of a galaxy, while on the other hand when tracing star-formation it is essential, as proven in the analysis, that the whole surface must be scanned.

On the whole, both methods, fiber or a standard long slit, ascertain that galaxies will have a different placement in the BPT diagrams and on certain occasions a different classification as more starlight is included in the aperture of spectral extraction. Starlight subtraction does not change significantly this effect, but it helps in some cases to measure emission lines, especially of higher ionization, that are otherwise completely lost in the stellar continuum. As demonstrated here reality appears to be more complex, with extra-nuclear star-forming activity (which cannot be subtracted) seriously influencing the photometry from large apertures. Low-activity galaxies (Low-luminosity AGN (LLAGN), Starburst, LINERs, and TO) seem to be more affected by star-forming activity throughout the galactic disk, often increasing their lower-ionization emission-line intensities and placing them higher in the activity classification diagrams. A larger sample of galaxies would prove beneficial in order to identify trends in the way galaxies are classified depending on the amount of host galaxy's light that is incorporated during their observations and furthermore examine the importance of their morphology by using as many different morphological types as possible.

\section*{ACKNOWLEDGMENTS}
This work was supported by a ``Maria Michail Manasaki" bequest fellowship. A. Maragkoudakis acknowledges partial support by HST grant AR-12621-01-A. A. Zezas acknowledges partial support by NASA grant NNX12AN05G, and Chandra grant AR1-12011X. We would like to thank the referee for his/her useful comments and suggestions which have improved the clarity of this paper. We also thank observers P. Berlind and M. Calkins for performing the FLWO observations. Funding for SDSS-III has been provided by the Alfred P. Sloan Foundation, the Participating Institutions, the National Science Foundation, and the U.S. Department of Energy Office of Science. The SDSS-III web site is http://www.sdss3.org/. SDSS-III is managed by the Astrophysical Research Consortium for the Participating Institutions of the SDSS-III Collaboration including the University of Arizona, the Brazilian Participation Group, Brookhaven National Laboratory, Carnegie Mellon University, University of Florida, the French Participation Group, the German Participation Group, Harvard University, the Instituto de Astrofisica de Canarias, the Michigan State/Notre Dame/JINA Participation Group, Johns Hopkins University, Lawrence Berkeley National Laboratory, Max Planck Institute for Astrophysics, Max Planck Institute for Extraterrestrial Physics, New Mexico State University, New York University, Ohio State University, Pennsylvania State University, University of Portsmouth, Princeton University, the Spanish Participation Group, University of Tokyo, University of Utah, Vanderbilt University, University of Virginia, University of Washington, and Yale University. This research has made use of the NASA/IPAC Extragalactic Database (NED) which is operated by the Jet Propulsion Laboratory, California Institute of Technology, under contract with the National Aeronautics and Space Administration.

\appendix
\section{Galaxy Characteristics}

\textbf{NGC 2608}: A barred spiral galaxy classified as star-forming based on the standard long-slit analysis of the nuclear spectrum. The galaxy changes its activity classification towards the TO region for intermediate aperture sizes and falls into the LINER region for the largest apertures in the long-slit approach. A large portion of the slit, oriented along the major axis, passes through an extensive interarm region. This could be responsible for the decrement in the flux of the Balmer lines (see the corresponding H$\alpha$ equivalent width plot), producing the upward and towards the right tendency on the BPT diagrams at small to intermediate sized apertures. Then, at larger apertures, the slit falls upon the spiral arm regions where we observe a gradual increase in the Balmer line flux and decrease of the BPT line ratios.

\textbf{NGC 2731}: A spiral star-forming galaxy with almost no scatter at the [O\,{\sc iii}]~$\lambda$5007/H$\beta$ axis but with significant variation towards the right side of the BPT diagrams mostly in the long-slit approach. As discussed previously, this is a case where the higher ionization  [O\,{\sc iii}]~$\lambda$5007 line incrementally decreases at faster rates compared to lower ionization lines as we move farther out from the nucleus of the galaxy. In the long-slit method the galaxy reaches the TO and LINER region at larger apertures.

\textbf{NGC 3306}: This is a barred-spiral galaxy and one of the starbursts in the sample having an upward movement on the BPT diagrams with increasing aperture. Several younger, extra-nuclear star-forming clumps on the spiral arms produce a harder ionizing continuum, which results in the upward behavior observed. Though close to the pure star-forming distinguishing line, it only relocates to the TO region for the largest apertures in the long-slit method.

\textbf{NGC 4412}: This is a typical example of a spiral Seyfert-2 galaxy having its AGN features obscured with increasing aperture and starlight. Regarding the change of its activity classification, the galaxy falls into the TO and star-forming regions in both methods and passes through the LINER region for intermediate apertures in the long-slit method.

\textbf{NGC 4491}: An early-type, barred spiral galaxy producing moderate-quality simulated elliptical spectra that have an upward trend on the BPT diagrams. No emission lines were measured without performing starlight subtraction. The two-dimensional spectrum doesn't show any evidence of extra-nuclear star-forming activity, and because the produced elliptical spectra were of moderate-quality it is possible that the applied starlight subtraction could have overestimated the flux on the absorption Balmer lines. This would give a smaller flux on the Balmer emission lines and would result in the observed upward behavior on the diagrams.

\textbf{NGC 4500}: Another early-type, barred spiral galaxy maintaining a TO classification in the [N\,{\sc ii}]~$\lambda$6583/H$\alpha$ diagnostic and a star-forming classification in the [S\,{\sc ii}]~$\lambda\lambda$6716, 6731/H$\alpha$, and [O\,{\sc i}]~$\lambda$6300/H$\alpha$ diagrams. The galaxy only passes into the LINERs' region at the largest apertures of the long-slit approach. Being in the TO region and very close to the maximum starburst line at its smallest aperture, the galaxy's possible AGN fraction appears to decrease with increasing aperture as in the case of Seyferts.

\textbf{NGC 4704}: This is a spiral Seyfert-2 galaxy showing a decrement of AGN characteristics with increasing aperture. The galaxy falls into the star-forming region passing through the TOs at intermediate apertures. Without starlight subtraction emission lines emerge approximately at half major semi-axis radius.

\textbf{NGC 4868}: An early-type spiral galaxy without notable scatter along all BPT axes, presenting no change of activity classification with increasing aperture. The galaxy's reddish color on the SDSS image and inspection of the two-dimensional spectrum suggests mild star-forming activity and agrees with the star-forming classification and placement towards the lower end of the [O\,{\sc iii}]~$\lambda$5007/H$\beta$ axis.

\textbf{NGC 5660}: Another starburst, late-type spiral galaxy presenting an upward movement on the BPTs due to extra-nuclear star-forming activity, as explained previously. Only the largest apertures cross the TO region in the [N\,{\sc ii}]~$\lambda$6583/H$\alpha$ diagnostic for the non-starlight-subtracted elliptical aperture and the standard long-slit methods. In the other two diagnostics the galaxy remains in the star-forming region.

\textbf{UGC 6732}: This is a lenticular galaxy classified as LINER based on the long-slit analysis of the nuclear spectrum. It shows a small upward tendency on the BPTs which could be statistically insignificant due to the uncertainties of the emission-line measurements. In the [S\,{\sc ii}]~$\lambda\lambda$6716, 6731/H$\alpha$, and [O\,{\sc i}]~$\lambda$6300/H$\alpha$ diagnostics, the galaxy crosses the Seyfert region at intermediate apertures. No emission lines were measured without starlight subtraction.

\textbf{UGC 9412}: This spiral Seyfert-2 galaxy is the second most distant galaxy in the sample. It moves towards lower values of [O\,{\sc iii}]~$\lambda$5007/H$\beta$ with increasing aperture, an effect diminished when applying starlight subtraction. Despite the decreasing [O\,{\sc iii}]~$\lambda$5007/H$\beta$ values, the galaxy preserves its Seyfert classification. Combined with the fact that it is relatively distant, the galaxy's AGN features seem to be most dominant and least affected by the host galaxy's starlight.

\textbf{3C 033}: This is an elliptical Seyfert galaxy having a small upward trend on the BPTs, but within the uncertainties of the emission-line measurements this could be statistically insignificant. Having a quiescent star-formation activity as an early-type galaxy, this small upward placement with increasing aperture is indicative of possible over-subtraction of the stellar continuum.

\textbf{3C 084}: An elliptical galaxy classified as LINER and producing the expected decrement of AGN features with increasing aperture in all three diagnostic diagrams.

\textbf{3C 296}: This elliptical galaxy is classified as Seyfert based on the analysis of the central elliptical aperture, descending towards the TO region in the [N\,{\sc ii}]~$\lambda$6583/H$\alpha$ diagnostic, and crossing the LINER region in the [O\,{\sc iii}]~$\lambda$5007/H$\beta$ diagnostic with increasing aperture. No emission lines were measured without starlight subtraction, and even so the $[S_{II}]\lambda\lambda 6716, 6731 $ doublet did not show any emission.

\end{document}